\documentclass[aps,nofootinbib,longbibliography,superscriptaddress,preprint]{revtex4-1}

\pdfoutput=1
\usepackage{amssymb, amsmath, graphicx}
\usepackage{color}
\usepackage{subfigure}

\DeclareMathOperator{\arcsinh}{arcsinh}
\DeclareMathOperator{\sgn}{sgn}

    \newcommand{\cL}{{\cal L}}

\newcommand{\cW}{{\cal W}}

\newcommand{\be}{\begin{equation}}
\newcommand{\ee}{\end{equation}}
\newcommand{\bea}{\begin{eqnarray}}
\newcommand{\eea}{\end{eqnarray}}

\newcommand\lrpar{\raise .8ex\hbox{$^\leftrightarrow$} \hspace{-9pt}
\partial}
\newcommand\lpar{\raise .8ex\hbox{$^\leftarrow$} \hspace{-9pt}
\partial}
\newcommand\rpar{\raise .8ex\hbox{$^\rightarrow$} \hspace{-9pt}
\partial}

\begin{document}
\title{Holographic dual of a time machine}

\author{Irina Arefeva}
  \email{arefeva@mi.ras.ru}
\affiliation{Steklov Mathematical Institute RAS, Moscow, Gubkina str. 8, 119991}

\author{Andrey Bagrov}
    \email{bagrov@lorentz.leidenuniv.nl}
\affiliation{Institute Lorentz, Leiden University, PO
  Box 9506, Leiden 2300 RA, The Netherlands}

\author{Petter S\"aterskog}
    \email{saterskog@lorentz.leidenuniv.nl}
\affiliation{Institute Lorentz, Leiden University, PO
  Box 9506, Leiden 2300 RA, The Netherlands}

\author{Koenraad Schalm}
    \email{kschalm@lorentz.leidenuniv.nl}
\affiliation{Institute Lorentz, Leiden University, PO
  Box 9506, Leiden 2300 RA, The Netherlands}

\begin{abstract}
We apply the $AdS/CFT$ holography to the simplest possible eternal time machine solution in $AdS_3$ based on two conical defects moving around
their center of mass along a circular orbit. Closed timelike curves in this space-time extend all the way to the boundary of $AdS_3$,
violating causality of the boundary field theory. By use of the geodesic approximation we address the ``grandfather paradox'' in the
dual $1+1$ dimensional field theory and calculate the two-point retarded Green function. It has a non-trivial analytical structure both at negative and
positive times, providing us with an intuition on how an interacting quantum field could behave once causality is broken.
In contrast with the previous considerations our calculations reveal the possibility of a consistent and controllable evolution of a quantum system without any need to impose additional consistency constraints.

\end{abstract}

\maketitle
\tableofcontents
\newpage
\section{Introduction}
Solutions to the equations of General Relativity that describe space-times containing closed timelike curves (CTC) have attracted significant interest
as they revealed at least hypothetical theoretical possibility of travelling in time. Since the renowned publication by Kurt G\"odel \cite{Godel:1949ga}
a number of causality violating solutions in GR as well as in modified theories of gravity have been constructed, among which we can name the
Tipler-Van Stockum time machine generated by axially rotating distribution of particles \cite{Van Stockum},
\cite{Tipler}, the Morris-Thorne-Yurtsever transversable wormhole \cite{MT,MTY}, the Gott time machine based on moving
 conical defects \cite{Gott}, the Ori dust solution \cite{Ori}, and the solutions in $f(R)$ theories of gravity \cite{f(R)} and theories with non-minimal
 matter-curvature coupling \cite{non-minimal}.

All questions about physics of time machines that could be posed in principle fall into three general categories:
\begin{itemize}
\item Is there a physical way to create a time machine?
\item Is there any time machine solution that can be stable?
\item What dynamical behaviour would a physical system experience evolving in a time machine background?
\end{itemize}
None of the questions have yet received a definite answer.

The answer to the first question is believed to be negative. Extensive analysis of particular time machine solutions has demonstrated that
in order to create a space-time with CTC one needs matter that violates strong, weak or null energy conditions of General Relativity (different solutions
require violation of different energy conditions), and only eternal
time machines can exist \cite{Carroll:1991nr, Deser:1991ye}. However we can not be sure that all matter in the
Universe obeys these conditions. For instance, there are a number of models of the dark energy violating the null energy condition \cite{Aref'eva:2006xy,Rubakov:2014jja,Creminelli:2010ba,Creminelli:2006xe}, and this provides a way to by-pass the no-go statement.

The second question was raised by Hawking in \cite{HawkingChronology}, where he conjectured that a space-time with CTC can be stable only on classical level, but will be unavoidably destroyed by quantum fluctuations of the metric. The real universal proof or refutal of the conjecture can be obtained only
within a framework of a complete theory of quantum gravity. String theory opened a possibility to check the chronology protection condition in specific cases. In
\cite{Costa} authors have shown that appearance of closed timelike curves in a certain (O-plane) orbifold background would cause
a Hagedorn transition that restructures the space-time transforming it into a chronologically safe configuration. So this result
can be considered as a very accurate and nice supporting evidence in favour of the Hawking conjecture.
On the other hand in \cite{Boyda:2002ba,Gimon:2004if,Brecher:2003rv}, it was demonstrated that the G\"odel type solutions can be smoothly embedded in the context
of string theory. Closed timelike curves in that case are hidden behind the so called holographic screens and do not violate causality in
the rest of the space-time. Thus the chronology is protected, but structure of the CTC remains unbroken by quantum effects.
An intriguing observation has been made by authors of \cite{AdSTM} and \cite{AdSTM2}, that from the point of view of the AdS/CFT correspondence, the existence
of CTC in the bulk can be related to negative or exceeding one fermionic probability in the boundary field theory.

The third category includes various types of the ``grandfather paradox''. For a classical wave equation on a non-globally hyperbolic space
with CTC the possibility of self-consistent dynamics was demonstrated in \cite{Arefeva:2009bz, VolovichGroshev}. Classical mechanical billiards and their self-consistency conditions have been studied in \cite{Echeverria:1991nk}.
A basis of states of a free quantum field theory in the Gott time machine has been constructed in \cite{Boulware}, where it was shown that the causality violation leads to an emergence of an effective non-unitary interaction in the theory. Non-unitarity of interacting field theories in time machines
was analyzed in \cite{Friedman:1992jc}.
Some authors even argued that evolution of a physical system along closed timelike curves can be studied experimentally by mean of
simulation of emergent gravity in metamaterials \cite{Metamaterials}, or a qubit interacting with an older version of itself \cite{Qubit}.

However the question about properties of an {\it interacting} quantum field theory in a time machine background remains open, though the real
``grandfather paradox'' can take place only in a self-interacting system.
When the notions of time ordering and unitarity are absent from the very beginning, it is unclear how to formulate
an interacting field theory. In this paper we address this problem and by use of the AdS/CFT correspondence provide a constructive
solution to it. Here we stand on the position that even if the presence of CTC causes breaking of unitarity in the boundary field theory \cite{AdSTM2},
it should not be regarded as a big problem as long as we can formulate a prescription how to solve the theory.
When one is trying to get an insight into
physics of paradoxical systems, it is not very useful to rely on the ``common sense'' intuition and corresponding fundamental
principles.

The AdS/CFT correspondence provides an elegant way to address the paradox.
In the large $N$ limit it relates quantum field theories to a classical gravity, and thus we
can study properties of a quantum theory in the CTC background just by careful analysis of the dual Riemannian geometry, without any need
to formulate special quantization rules that would be valid in the case of broken causality.

The paper is organized as follows. In the next section we introduce a simple set up for the time machine in $AdS_3$.
In Sec. \ref{sec:Entwinement} we discuss geodesic structure of the spacetime, and suggest that it could lead to non-trivial effects in the boundary
field theory. In Sec. \ref{sec:Quasi} we introduce a notion of timelike quasigeodesics that will be then used for connecting timelike separated
boundary points. Finally, in Sec. \ref{sec:Green} we formulate a precise algorithm for the Green function evaluation, provide the results of numerical
simulations, and discuss the related phenomenology.

\section{Time machine in $AdS_3$ \label{sec:TM}}
The eternal time machine solution in $AdS$ has been suggested by Gott and DeDeo in \cite{GottDeDeo} (for similar solutions containing
CTC but collapsing into a BTZ black hole see \cite{Matschull}). Here we briefly recall its structure closely following the original text.

The three dimensional global anti-de Sitter space-time can be thought of as a hypersurface
\begin{equation}
\label{surf}
-X_0^2-X_3^2+X_1^2+X_2^2=-1,
\end{equation}
embedded in a four-dimensional flat ${\mathbb R}^{2,2}$ space-time with a metric:
\begin{equation}
\label{met}
ds^2=-dX_0^2-dX_3^2+dX_1^2+dX_2^2.
\end{equation}
In the Schwarzschild coordinates the embedding formulas are
\bea \label{eq:SchwarzschildGlobal}
X_0 & = & \sqrt{1+R^2}\cos t \,, \\
X_3 & = & \sqrt{1+R^2}\sin t \,, \nonumber\\
X_1 & = & R\cos\phi\,, \nonumber \\
X_2 & = & R\sin\phi\,,\nonumber
\eea
where $R\in(0,\infty),t \in(-\infty,\infty),\,\phi\in[0,2\pi)$.

The induced metric is then
\be ds^2=-(1+R^2)dt ^2+\frac{dR^2}{1+R^2}+R^2d\phi^2.\ee
\begin{figure}
\begin{center}
\begin{tabular}{cc}
\subfigure[]{\includegraphics[width=0.5\textwidth]{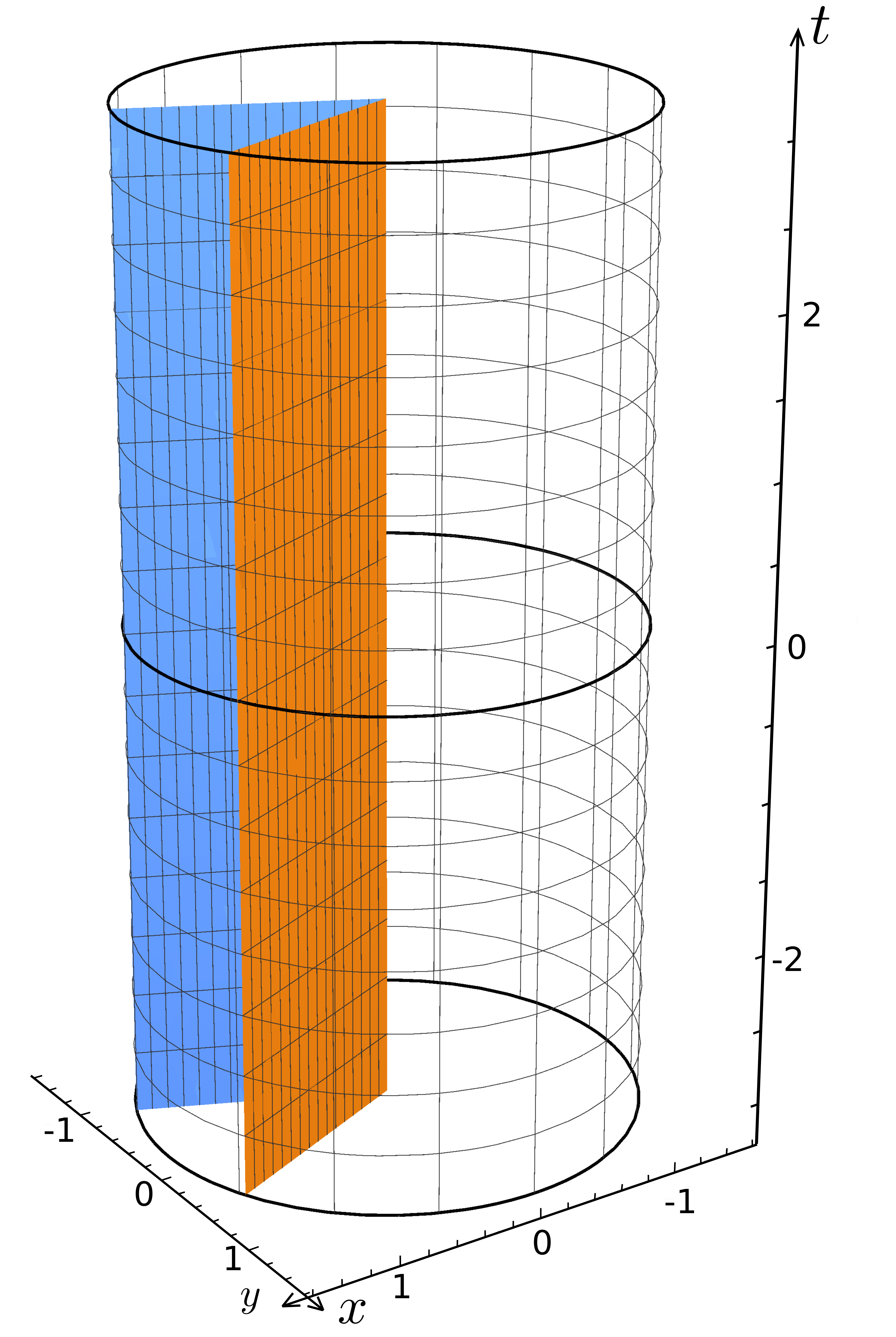}}
\subfigure[]{\includegraphics[width=0.5\textwidth]{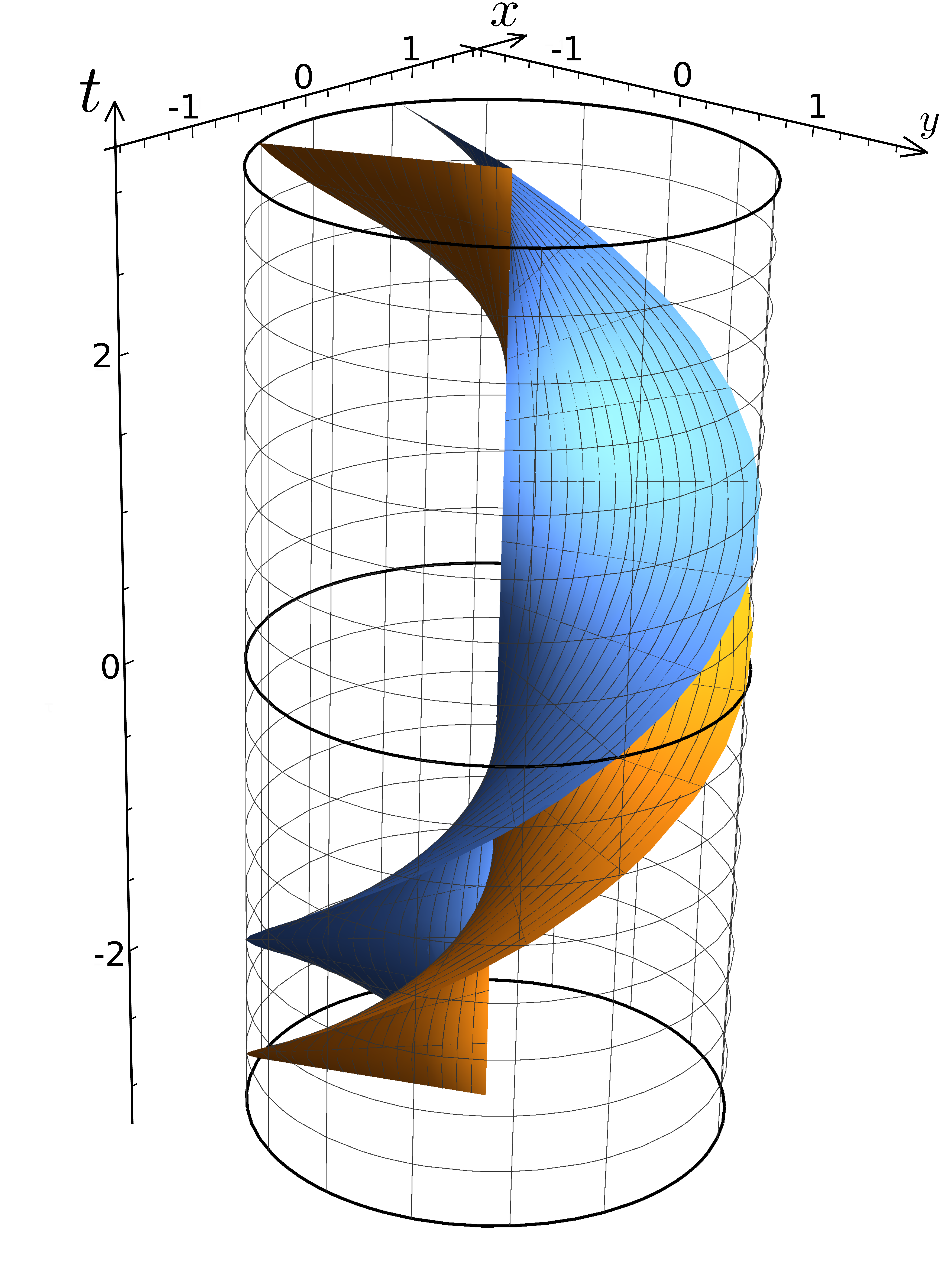}}
\end{tabular}
\caption{Two equivalent unfoldings of the $AdS_3$ spacetime with a conical defect $\alpha=\sqrt{3}\pi$. The larger part of the spacetime outside of the faces is to be
cut out, and the faces are identified. The only physical space is the narrow region between the faces. To construct a time machine the twisted unfolding is more convenient to use.}
\label{fig:SingleCone}
\end{center}
\end{figure}
A massive particle put into a three dimensional space-time removes a wedge with an angle deficit proportional to the
mass of the particle, and edges (faces) emerging from this point-like particle. Points on the opposite edges of the wedge are identified, and the
resulting space-time contains a conical defect, Fig.\ref{fig:SingleCone}(a). When we are looking at the unfolding of the conical defect, coordinate locations of the edges
do not have an independent physical meaning, and we are free to rotate them preserving the angular deficit.
For our purposes it will be convenient to make the cut out ``pizza slice'' twist in time with a constant angular velocity in the
reference frame of the massive particle, making a full rotation in a period $2\pi$, Fig.\ref{fig:SingleCone}(b).
Then for the trailing and leading faces of the wedge in the embedding coordinates we get:
\be
\begin{matrix}\label{eq:SchwarzschildGlobalEdges}
X_0^{t} &=& \sqrt{1+R^2}\cos t  & X_0^{l} = \sqrt{1+R^2}\cos t \,, \\
X_3^{t} &=& \sqrt{1+R^2}\sin t   & X_3^{l} = \sqrt{1+R^2}\sin t \,,\nonumber \\
X_1^{t} &=& R\cos(t -\alpha/2)  & X_1^{l} = R\cos(t +\alpha/2)\,,\nonumber \\
X_2^{t} &=& R\sin(t -\alpha/2)  & X_2^{l} = R\sin(t +\alpha/2)\,. \nonumber
\end{matrix}
\ee
Here $\alpha$ is the angular deficit of the conical spacetime. Integrating the spacetime stress-energy tensor
 over the angle, we can deduce that the effective mass concentrated in the interior of the bulk is
\be M=-\frac{1}{8G}+\frac{\alpha}{16\pi G} .\ee
Here the second term is the mass of the point-like source, and the first one is the contribution from the negative $AdS$ curvature.

For a single static conical defect we can not make its angular deficit $\alpha$ larger than $2\pi$.
If the mass of the point-like particle exceeds the limit $\alpha=2\pi$ ($M\geq 0$),
the resulting space-time will be rather a BTZ black hole instead of a naked conical singularity \cite{Balasubramanian:1999zv}.
\begin{figure}
\begin{center}
\begin{tabular}{cc}
\subfigure[]{\includegraphics[width=0.5\textwidth]{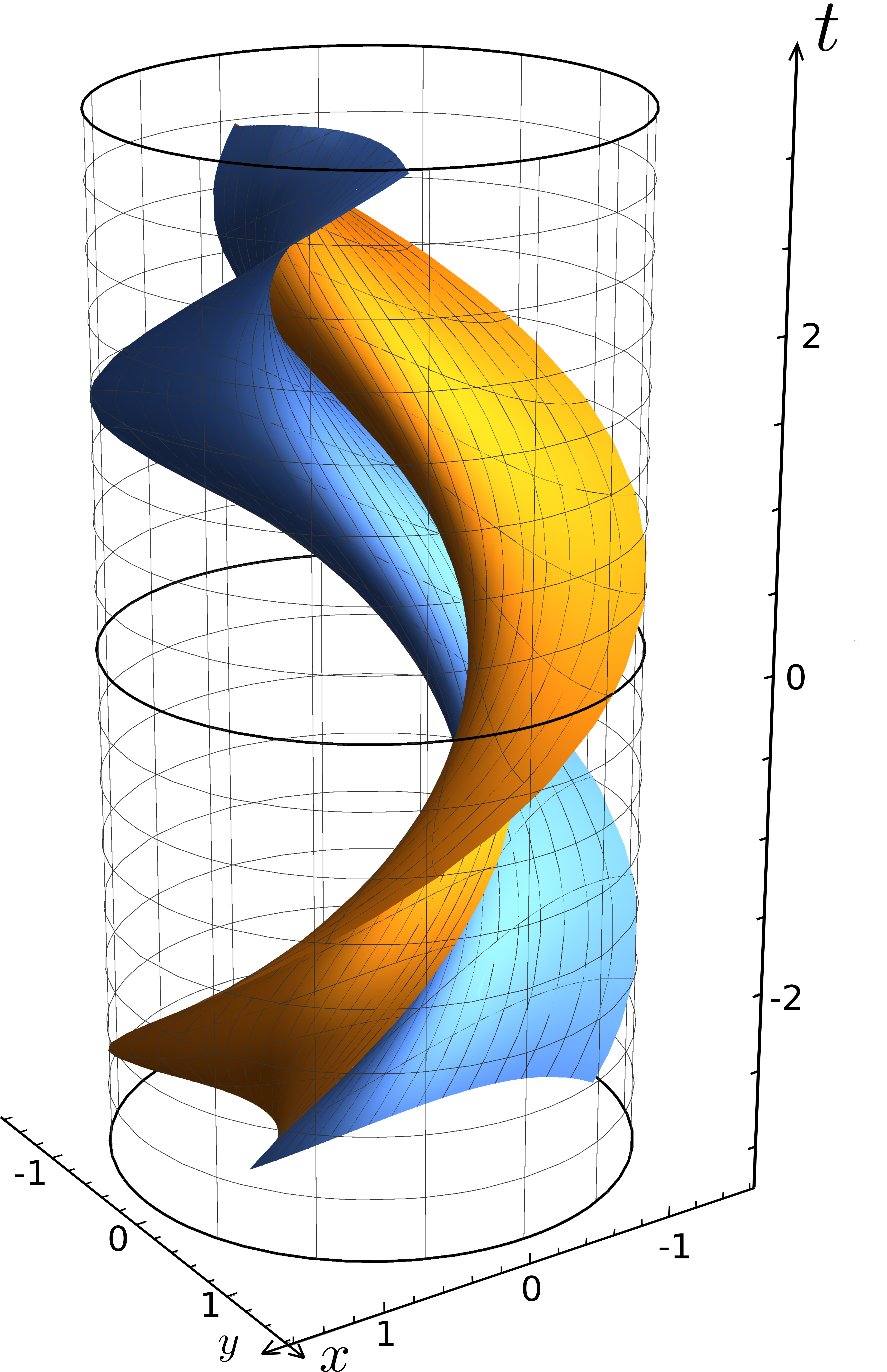}}
\subfigure[]{\includegraphics[width=0.5\textwidth]{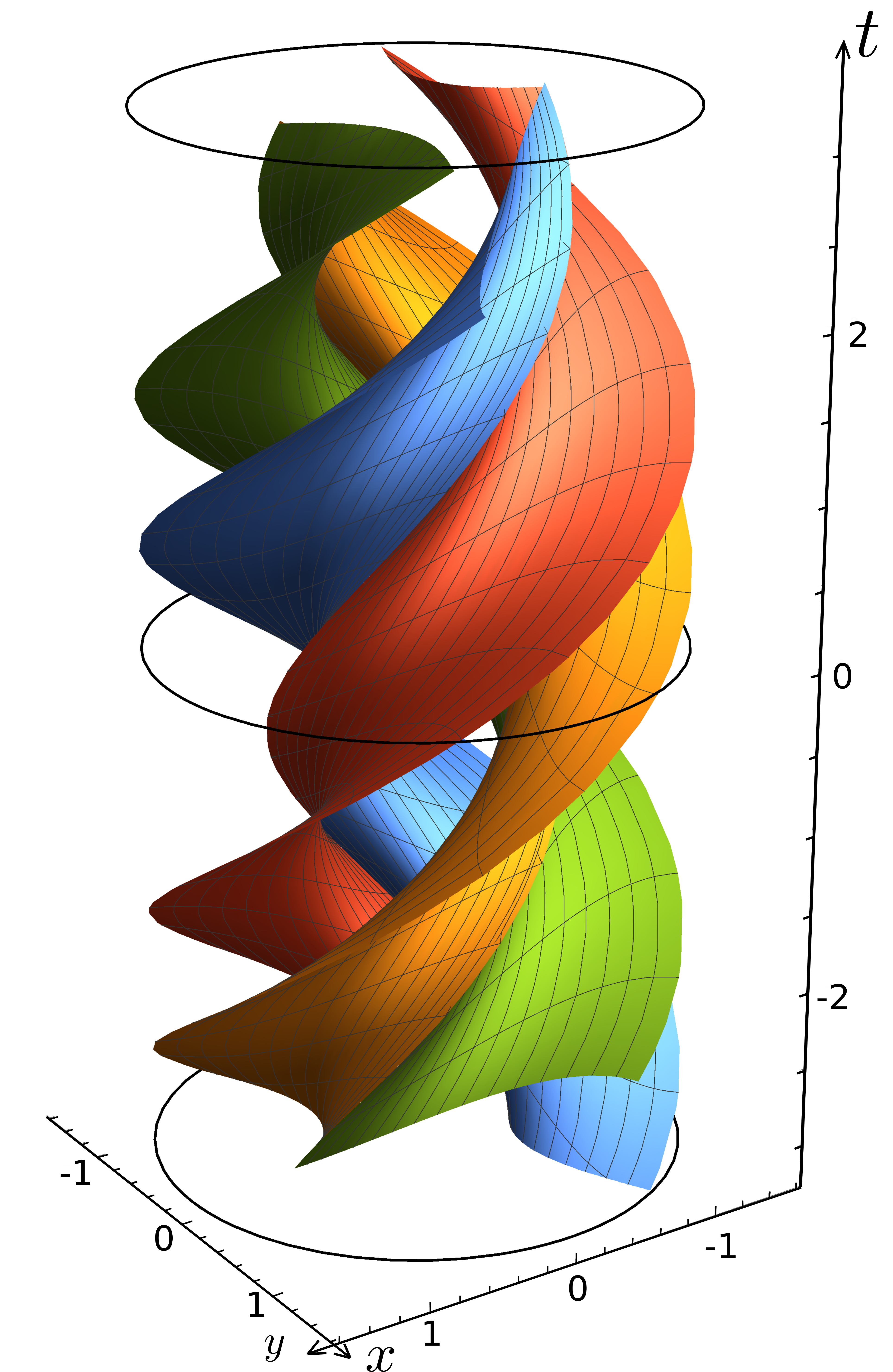}}
\end{tabular}
\caption{(a) A boosted conical defect in $AdS_3$. Faces of the wedge are deformed, and the identification occurs between
points with different time coordinates (in the centre of mass frame). Here $\alpha=\sqrt{3}\pi$, $\psi=1$. (b) The DeDeo-Gott time machine.}
\label{fig:BoostedCone}
\end{center}
\end{figure}

However, if we boost the massive source, the removed wedge is effectively getting ``squeezed'' from the point of view of
an external observer at rest (see Fig. \ref{fig:BoostedCone}(a)). This provides a room for a second conical defect with a deficit angle $\beta$
such that $\alpha+\beta>2\pi$.
In other words, relative motion can support
 the system of conical defects, preventing them from collapsing into a black hole. This will be the essence of the DeDeo-Gott construction.

Consider a system of two identical conical defects undergone two opposite Lorentz transformations, in the $(X_0,X_1)$ and $(X_3,X_2)$ planes of the embedding
space each:
\be
\Lambda_I=\Lambda_{II}^{-1}=
\left( \begin{matrix}
\cosh\psi & 0 & \sinh\psi & 0 \\
0 & \cosh\psi & 0 & \sinh\psi \\
\sinh\psi & 0 & \cosh\psi & 0 \\
0 & \sinh\psi & 0 & \cosh\psi
\end{matrix}\right) \label{eq:EmbedLorentz}
\ee
In the three-dimensional coordinates of the $AdS$ spacetime these Lorentzian transformations correspond to $SO(2,2)$ isometry
transformations.

It can be shown that in the coordinates of global $AdS_3$ these defects move along the same circular orbit $R=const$ with a
constant velocity, always being at the opposite points of the orbit, Fig.\ref{fig:BoostedCone}(b).
A conical defect sits at $\tilde{R}=0$ in its rest frame, i.e.
\be \tilde{X}_0=\cos t,\,\,\tilde{X}_1=0,\,\,\tilde{X}_2=0,\,\,\tilde{X}_3=\sin t\,.\ee
In the boosted frame
\be X_0=\cosh \psi \cos t,\,\,X_1=\sinh \psi \cos t,\,\,X_2=\sinh \psi \sin t,\,\,X_3=\cosh \psi \sin t\,,\ee
hence
\bea R&=&\sqrt{X_1^2+X_2^2}=|\sinh \psi|,\\ \cos \phi&=& \frac{X_2}{R}=\frac{\sinh \psi \cos t}{|\sinh \psi|}=\sgn \psi \cdot \cos t\,.\eea

In the rest frame of a wedge, the points on its edges are identified at equal coordinate times. However if we boost it,
from the point of view of an external observer this identification would occur at different times leading to time jumps for a particle moving around
the conical singularity. In the case of a single conical defect its boost can be regarded as a global coordinate transformation of the space-time,
which obviously can not cause any new physical effects. However, with two defects moving relative to each other, the relative time jumps
become a physical effect that can not be eliminated by a (proper) choice of coordinate system. These time jumps allow for the existence of CTC.

\begin{figure}
\begin{center}
\includegraphics[width=0.5\textwidth]{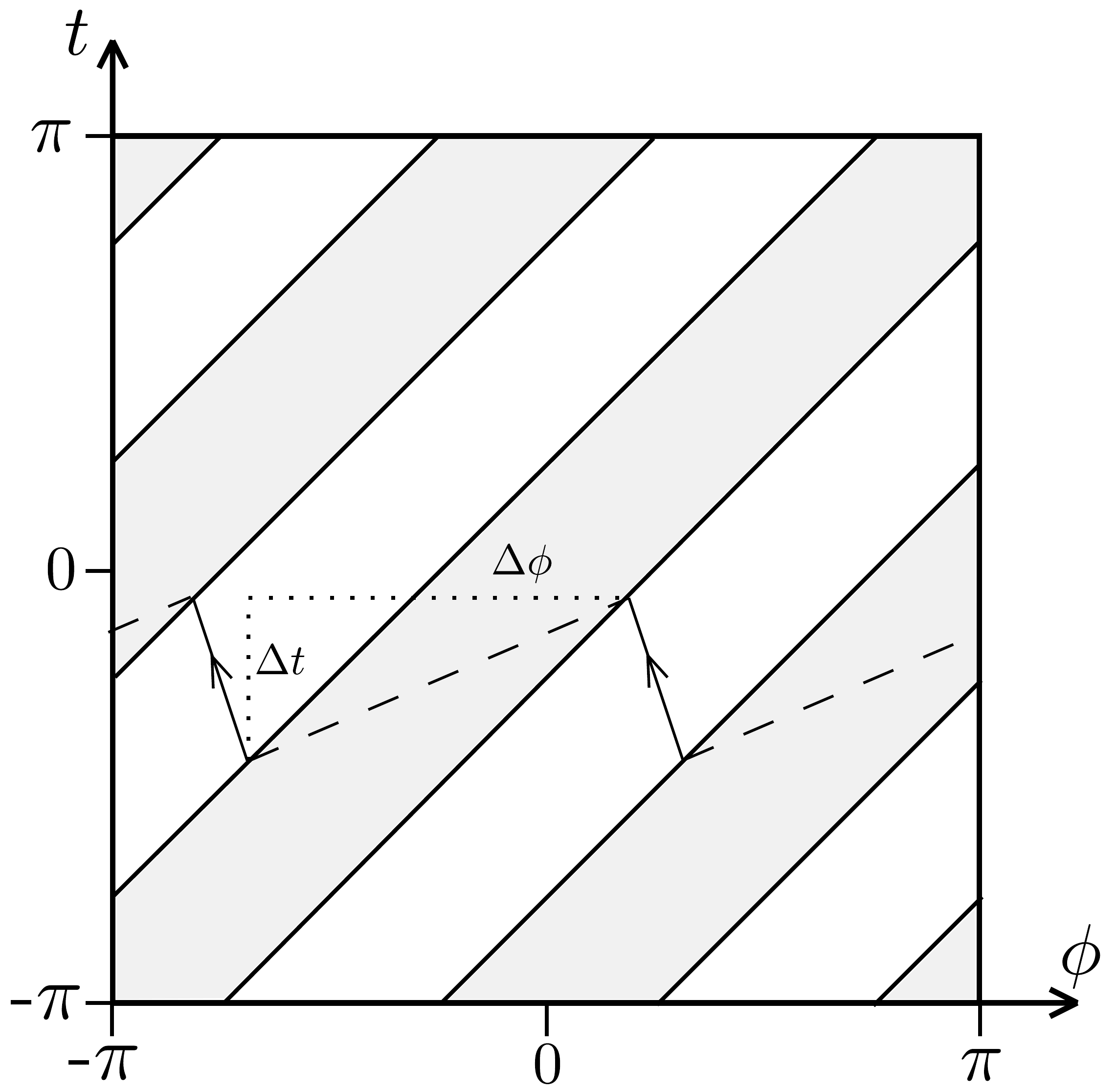}
\caption{A timelike particle moving along the boundary experiences a jump in time and angle when it hits a face of one of the wedges. If $\Delta t+\Delta\phi>\pi$,
closed timelike curves become possible. The gray strips are the cut out part of the boundary, and the white strips are the physical part of the boundary.}
\label{fig:BoundaryStrip}
\end{center}
\end{figure}

Existence of closed timelike curves in this space-time can be demonstrated by looking at the identification of the edges near the
boundary of $AdS_3$ (at $R\rightarrow \infty$). We refer the reader to \cite{GottDeDeo} for a detailed discussion, here we just quote the result.
Speaking in terms of the unfolding of the two-conical space-time, when a timelike particle living on the boundary of
the $AdS$ cylinder hits an edge of one of the two wedges, it undergoes a time and an angle jump:
\bea
\Delta t  & = & 2\arctan\left(\frac{\sin(\alpha/2)\tanh\psi}{1+\cos(\alpha/2)\tanh\psi}\right)\,, \label{CTC}\\
\Delta\phi & = & 2\arctan\left(\frac{\sin(\alpha/2)}{\tanh\psi+\cos(\alpha/2)}\right)\,.
\eea
It can be shown that $\Delta t +\Delta\phi=\alpha$. If $\alpha\geq\pi$, the world line of the particle becomes a closed timelike curve,
and thus the space-time is a time-machine, see Fig.\ref{fig:BoundaryStrip}.

In the regime when this limit is not exceeded, and CTC are not present, the space-time has been studied in detail from holographic point
of view in \cite{Balasubramanian:2003kq}, but the case of broken causality has not been addressed.

In the next sections we will study geodesic structure of this time machine and explicitly show that in presence of the two orbiting conical defects
we deal with a highly-nontrivial lensing of geodesics, and this reflects on the structure of two point Green's functions of the dual boundary theory.

\section{Entwinement of geodesics and causality violations \label{sec:Entwinement}}
When conformal dimension $\Delta$ of a boundary operator in $AdS/CFT$ is very high, the corresponding two-point Green function can be derived in the geodesic
approximation \cite{Balasubramanian:1999zv}:
\be G(A,B)=e^{-\Delta \cL_{AB}}\,,\ee
where $\cL_{AB}$ is the length of a geodesic connecting boundary points $A$ and $B$. If there are more than one geodesic between $A$ and $B$, they can give additional contributions to the propagator.
This is the case for the DeDeo-Gott time machine geometry, and here we address possible outcome of this in details.

\begin{figure}
\begin{center}
\begin{tabular}{cc}
\subfigure[]{\includegraphics[width=0.5\textwidth]{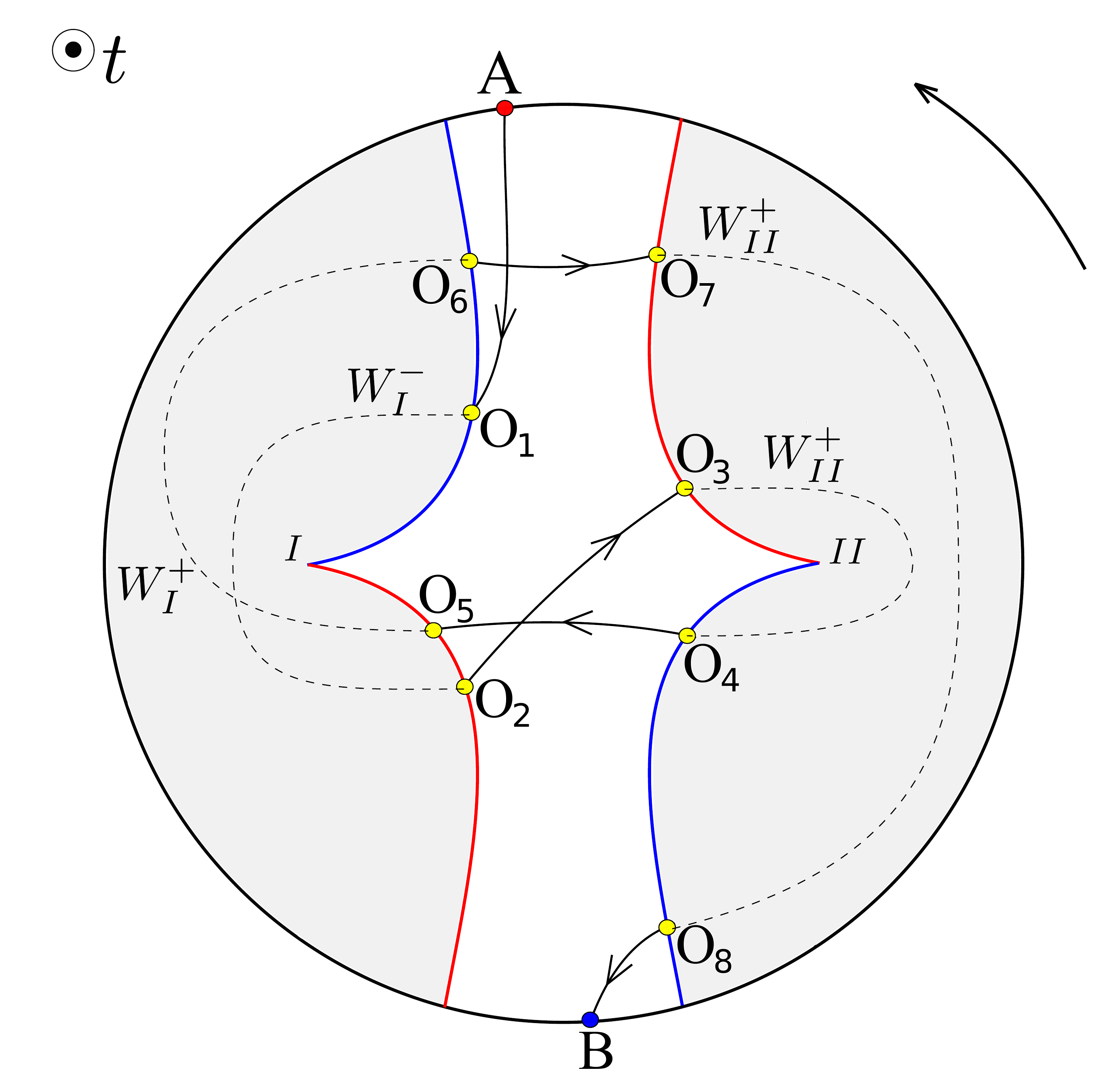}}
\subfigure[]{\includegraphics[width=0.5\textwidth]{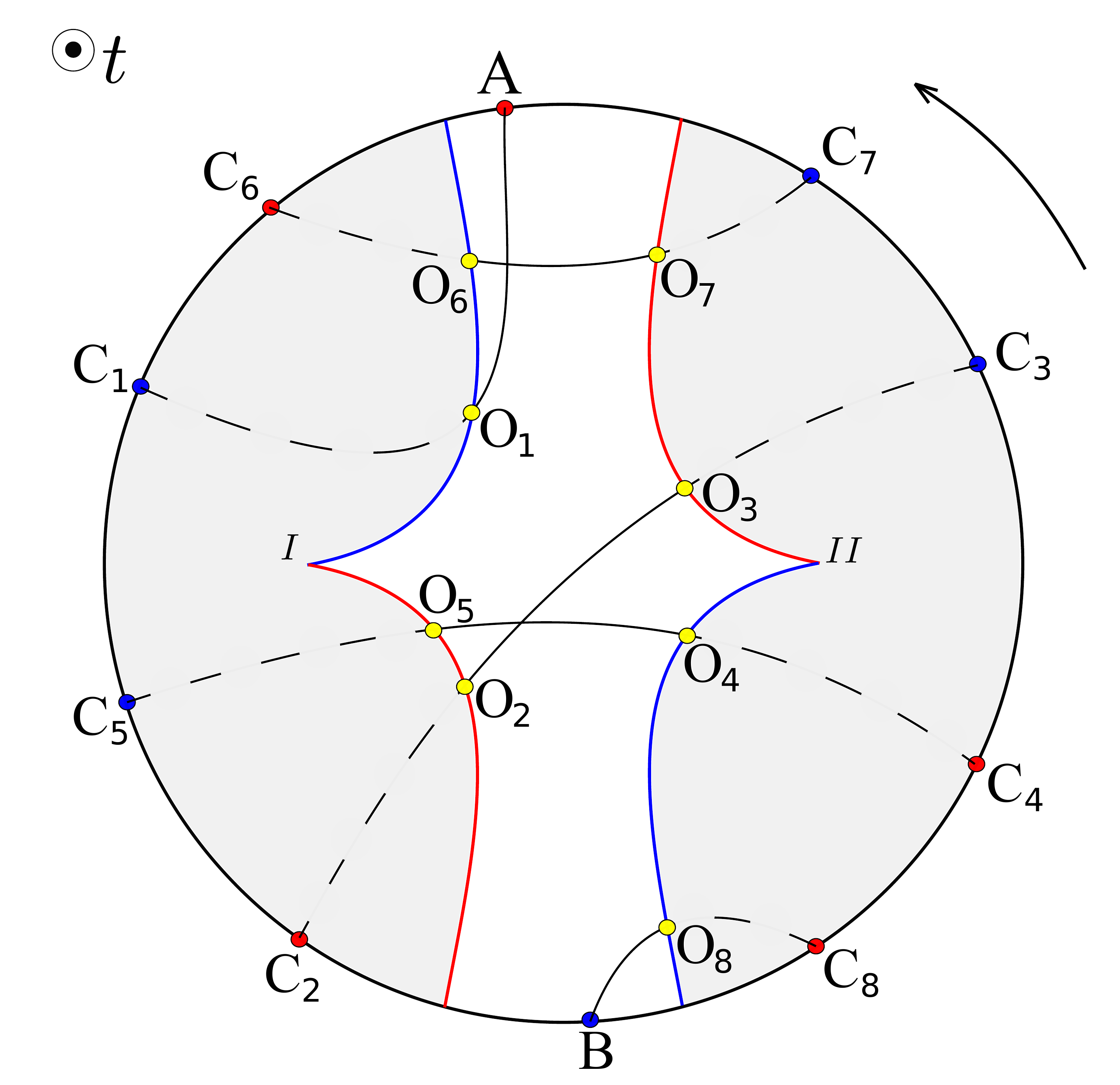}}
\end{tabular}
\caption{Schematic constant time projection of a typical geodesic connecting points A and B in the time machine. Red curves are for leading faces of the
rotating wedges, and blue curves - for trailing faces. The entwinement configuration in this particular case is
${\cal W}^-_I {\cal W}^+_{II} {\cal W}^+_I {\cal W}^+_{II}$ according to the notations introduced in the main text.
All shown points in principle can have different time coordinates (here we schematically project them down to a single time section, so the curves the geodesic is made from
should be taken only as an approximate artistic representation). On picture (a) the acts of entwinement and identifications are shown explicitly.
$A$ and $B$ belong to the physical unremoved part of the spacetime, and $O_i$ are the points where the geodesic undergoes ``refraction'' on the wedges.
Picture (b) demonstrates the idea of complementary points $C_i$ located in the removed part of the spacetime.
}
\label{fig:TypicalWinding}
\end{center}
\end{figure}

Consider two arbitrary points $A\left(t _1,\phi_1\right)$ and $B\left(t _2,\phi_2\right)$ located in the physical (unremoved) part of the $AdS_3$ boundary.
Having two rotating conical defects in the bulk makes the structure of possible geodesics connecting\footnote{When $A$ and $B$ are timelike separated
we encounter some subtleties caused by the fact that in the $AdS$ space-time a timelike geodesic can not reach the boundary. These issues
will be commented further on, but the general point of view described in this section remains unchanged.} $A$ and $B$ very nontrivial, so we should find a way
to calculate their contributions to the two-pont Green's function $G(A,B)$. Let us shoot a geodesic from the boundary point $A$ to
the point $B$. Before it hits the point $B$ it can undergo a number of ``refractions'' on the faces of wedges, winding around either of two conical defects clockwise
(if it hits the leading face of the wedge head on) or counterclockwise (if it overtakes the trailing face of the wedge from behind).
For example, schematically a typical geodesic may have a structure (see also Fig. \ref{fig:TypicalWinding})
\begin{equation}
 A \rightarrow {\cal W}^-_{I} \rightarrow {\cal W}^+_{II} \rightarrow {\cal W}^+_{I} \rightarrow {\cal W}^+_{II} \rightarrow B,
\end{equation}
where ${\cal W}^{+,-}_{I,II}$ stands for the act of clockwise/counterclockwise winding around the 1-st or the 2-nd wedge respectively.

So, formally the Green's function in the geodesic approximation is given by
\begin{equation}
 G(A,B)=\sum\limits_{n=0}^{\infty}\sum\limits_{\left\{ {\cal W}_{1} \cdots {\cal W}_{n} \right\}}e^{-\Delta \cL(A\left\{ {\cal W}_{1} \cdots {\cal W}_{n} \right\}B)},
 \label{genGreen}
\end{equation}
where the second sum is taken over all different entwinement structures corresponding to the same number
of windings, and the first sum is taken over all winding numbers\footnote{This idea of entwinements in holography has been introduced in \cite{Balasubramanian:2014sra}, but there it was related to a concept of entanglement entropy ``shadows'' rather than to subleading contributions
to the propagator}.

It is easy to see that for a given number of windings $N$ the maximal possible number of topologically different geodesics is
\be n_N = 4\cdot 3^{N-1}\,.\ee
The first winding act can be of four different types. But for each of the next steps, if a geodesic wrapped around a conical defect,
for example, clockwise, then on the next step it can not go in the opposite direction and
 wrap around the same conical defect counterclockwise. It means that in the sequence of windings
the winding act $\cW^+_I$ can be followed (at least hypothetically) by $\cW^+_{I},\,\cW^+_{II},\,\cW^-_{II}$, but not by
$\cW^-_I$.

As we will see further, for a given pair of boundary points $(A,B)$ not all sequences of entwinements are physically realistic and can contribute to the sum
\eqref{genGreen}.

To find a proper prescription for the lengths of the non-trivial winding geodesics let us discuss in detail a particular example.

\begin{figure}
\begin{center}
\begin{tabular}{cc}
\subfigure[]{\includegraphics[width=0.5\textwidth]{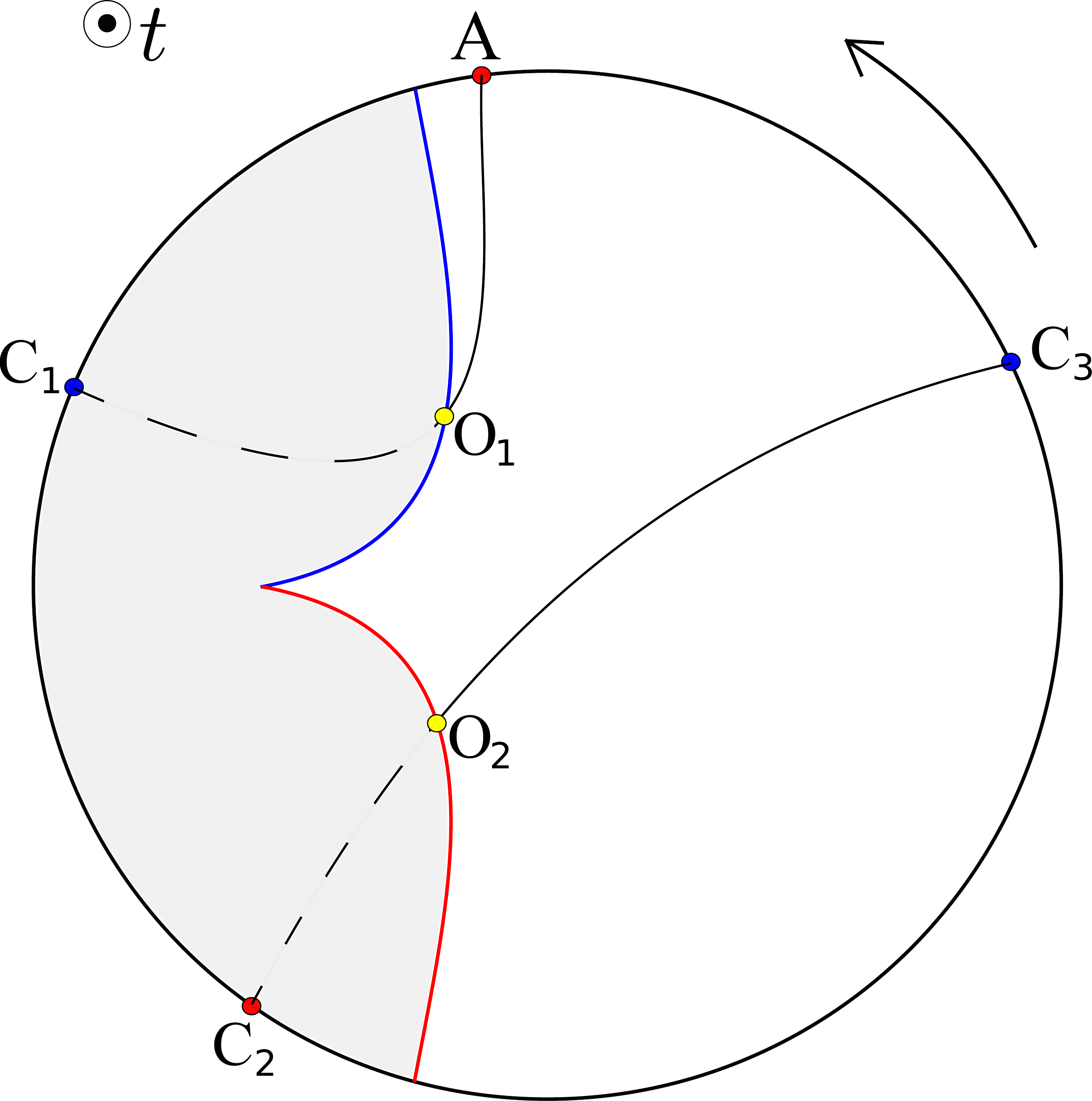}}
\subfigure[]{\includegraphics[width=0.5\textwidth]{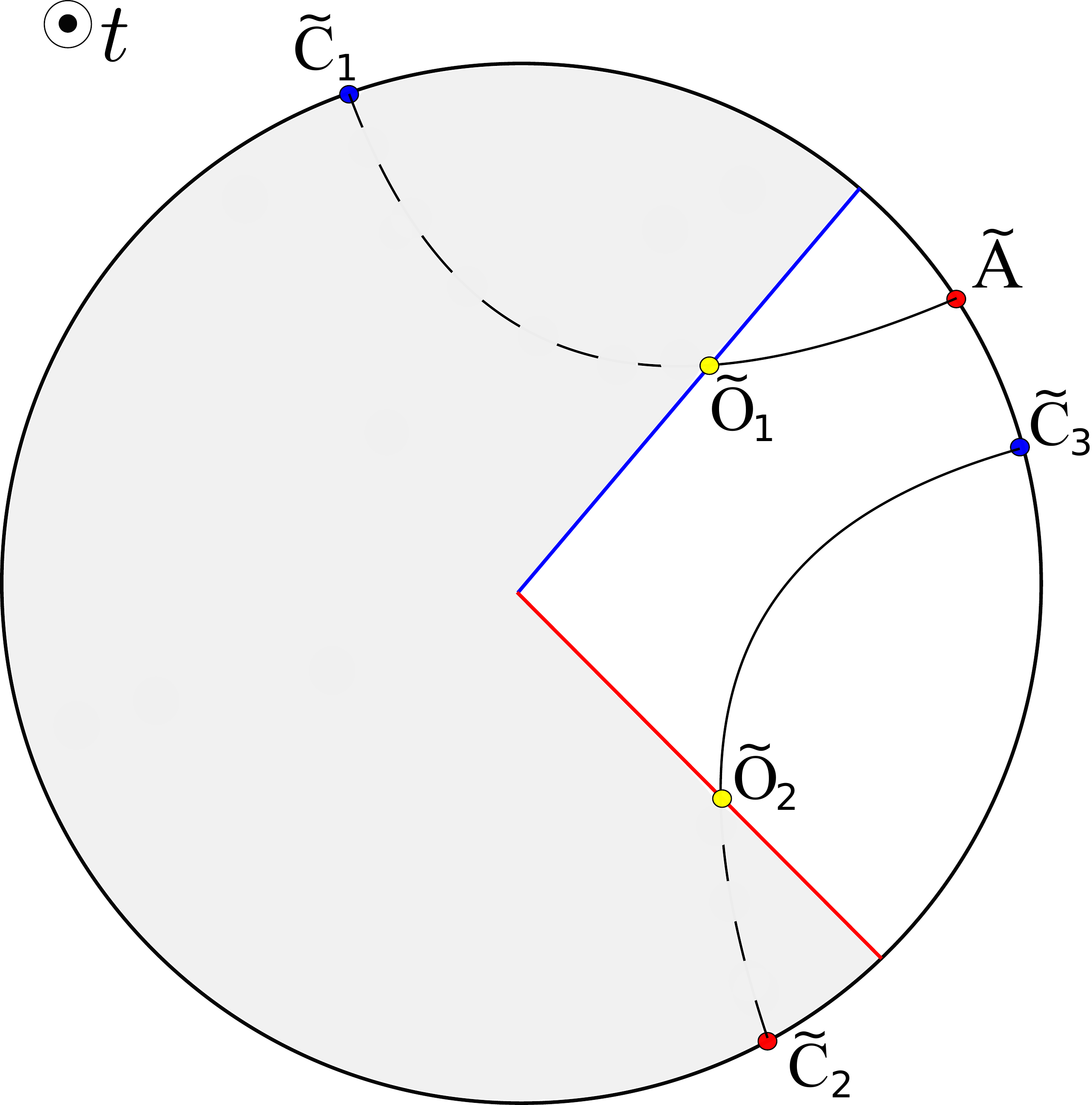}}
\end{tabular}
\caption{The left picture demonstrates ${\cal W}^-_I$ refraction of a geodesic on the first conical defect.
The right picture demonstrates how it looks like in a reference
frame of the conical defect. Again, points generically belong to different time slices, and the representation is purely schematic.}
\label{fig:BoostUnboost}
\end{center}
\end{figure}

Take a look at Fig.\ref{fig:TypicalWinding}. The length of the presented geodesic is a sum of lengths of its' composing arcs:
\begin{equation}
 \cL(A \rightarrow {\cal W}^-_I {\cal W}^+_{II} {\cal W}^+_{I} {\cal W}^+_{II} \rightarrow B)=
 \cL_{AO_1}+\cL_{O_2O_3}+\cL_{O_4O_5}+\cL_{O_6O_7}+\cL_{O_8 B}. \label{eq:SumOfArcs}
\end{equation}
We will refer to points $C_i$ as complementary points, and $O_i$ as refraction points.
Coordinates of the refraction points $O_i$ are to be found from coordinates of $A$ and $B$, and that can be easily done step by step.

Focus on the first refraction on the wedge, ${\cal W}^{-}_{I}$. The points of interest are $\left\{A,\, O_1,\,C_1,\,C_2,\, O_2,\,C_3\right\}$. Here we neglect for a while
the second wedge, so we do not consider the point $O_3$ at all, and we treat $C_3$ as a physical point (not just as a point in the complementary ``removed'' space),
see Fig.\ref{fig:BoostUnboost}(a). These six points can be regarded as a result of boost transformation $\Lambda_I$ applied to the wedge. We can ``unboost'' the wedge and find static pre-images of
these points (see Fig.\ref{fig:BoostUnboost}(b)). For the ``unboosted'' points the following relations trivially hold
\begin{align}
 \widetilde{C}_2=\Lambda_I^{-1}C_2=\Lambda_I^{-1}A-(0,\alpha)=\widetilde{A}-(0,\alpha),\\
 \widetilde{C}_3=\Lambda_I^{-1}C_3=\Lambda_I^{-1}C_1-(0,\alpha)=\widetilde{C}_1-(0,\alpha),\nonumber
\end{align}
where $(0,\alpha)$ is a boundary identification vector proportional to the angular deficit. Here we subtract the identification vector $(0,\alpha)$ because this particular entwinement is counterclockwise. For clockwise ${\cal W}^+$ we should rather add $(0,+\alpha)$.
In that case points $\widetilde{O}_1$ and $\widetilde{O}_2$ are intersections of geodesics $\widetilde{A}\widetilde{C}_1$ and $\widetilde{C}_2\widetilde{C}_3$
with faces of the static wedge.

Applying the same procedure to the other entwinements, in a generic case we get a system of recurrent relations
\bea
C_2&=&\Lambda_1 (\Lambda_1^{-1}A\pm(0,\alpha)), \label{eq:recurrent} \\
C_{2j}&=&\Lambda_j (\Lambda_j^{-1}C_{2j-2}\pm(0,\alpha)),\nonumber\\
C_{2N-1}&=&\Lambda_N (\Lambda_N^{-1}B\mp(0,\alpha)),\nonumber\\
C_{2j-1}&=&\Lambda_j (\Lambda_j^{-1}C_{2j+1}\mp(0,\alpha)).\nonumber
\eea
Here $\Lambda_j=\Lambda_I$ if the corresponding winding is ${\cal W}^{\pm}_I$.
$\Lambda_j=\Lambda_{II}$ if the corresponding winding is ${\cal W}^{\pm}_{II}$.
In these formulae we pick up the upper sign if ${\cal W}^{+}_{I,II}$,
and the lower sign if ${\cal W}^{-}_{I,II}$.

Note, that the Lorentz boost we have defined in terms of the embedding space coordinates acts non-linearly on the $AdS_3$ points,
therefore we can not simply expand the parentheses in \eqref{eq:recurrent}.

Then for each of the auxiliary arcs $C_{2k}C_{2k+1}$ we can derive coordinates of the refraction points $O_{2k},\,O_{2k+1}$,
and write down corresponding lengths of the composing arcs.

Later we will also show that not every formally generated sequence of windings does exist.

\section{Quasigeodesics connecting boundary points \label{sec:Quasi}}
To discuss causality properties of the dual boundary QFT, we will in particular need to consider boundary points with timelike separation.
The conceptual problem we unavoidably encounter here is the absence of timelike geodesics connecting points on the
conformal boundary of $AdS$. The equations for timelike geodesics can be derived from the following Lagrangian for a massive particle in $AdS$:
\be -(1+R^2)\dot{t}^2+\frac{\dot{R}^2}{1+R^2}+R^2\dot{\phi}^2=-1 \,.\ee
Such a particle has two conserved momenta:
\bea {\cal E} &=& (1+R^2)\dot{t},\\
{\cal J} &=& R^2\dot{\phi}.
\eea
Substituting them into the Lagrangian we obtain the radial equation of motion:
\bea -\frac{{\cal E}^2}{1+R^2}+\frac{\dot{R}^2}{1+R^2}+\frac{{\cal J}^2}{R^2}=-1,\\
\dot{R}^2=-\left(1+R^2\right)\left(1+\frac{{\cal J}^2}{R^2}\right)+{\cal E}^2\,.\eea
Clearly the right hand side of the equation turns negative as $R\rightarrow \infty$, and thus
no real solution to this equation can exist.

In the case of a stationary spacetime the obstacle could be easily surmounted by performing the analytic continuation of the metric to the
Euclidean signature, calculating the Green function in terms of Euclidean lengths of the geodesics, and making the
inverse Wick rotation back to real time. However in our case we deal with a spacetime that is not only non-stationary, but wich has no
good global notion of time. Hence we are forced to stick to the Lorentzian time.

The way to implement the geodesic approximation for timelike separated boundary
points in the single Poincar\'e patch has been suggested in \cite{Balasubramanian:2012tu}. Let's turn for a second to the
single patch of the $AdS_3$ spacetime, covered by the Poincar\'e coordinates:
\be ds^2=-r^2dt^2+\frac{dr^2}{r^2}+r^2dx^2\,.\ee
Again, a massive bulk particle has two kinetic invariants:
\bea  E = r^2\dot{t},\\
J = r^2\dot{x}\,,
\eea
but now we have two different classes of spacelike geodesics.
\begin{itemize}
\item For $J^{2}>E^{2}$:
\begin{equation}
\begin{cases}
r(\lambda)=\sqrt{J^{2}-E^{2}}\cosh\lambda \\
x (\lambda)=x_{0}+\frac{J}{J^{2}-E^{2}}\tanh\lambda \\
t(\lambda)=t_{0}+\frac{E}{J^{2}-E^{2}}\tanh\lambda
\end{cases}
\end{equation}
\item For $E^{2}>J^{2}$:
\begin{equation}
\begin{cases}
r(\lambda)=\sqrt{E^{2}-J^{2}}\sinh\lambda \\ \label{eq:Belgian}
x(\lambda)=x_{0}-\frac{J}{E^{2}-J^{2}}\coth\lambda \\
t(\lambda)=t_{0}-\frac{E}{E^{2}-J^{2}}\coth\lambda
\end{cases}
\end{equation}
\end{itemize}
We will be interested in the geodesics of the second kind. As $\lambda=0$
these geodesics approach the point $r(0)=0$, which is the Poincar\'e horizon of the
half-$AdS$ chart. Regarding the horizon as a single infinitely far point (as in the theory of complex functions),
we can consider two disconnected spacelike geodesics possessing the same kinetic invariants $E$ and $J$, but emerging from two different timelike
separated boundary points $A(t_A,x_A)$ and $B(t_B,x_B)$, as two branches of a single geodesic reaching the spatial infinity and returning back to the boundary. The length of
such a geodesic will be divergent not only as $r\rightarrow \infty$ (the standard holographic UV divergence), but also as $r\rightarrow 0$,
but this can be cured by an appropriate
renormalization\footnote{For details see App. B and D of \cite{Balasubramanian:2012tu}}. The resulting expression for the renormalized length of the geodesic is simply
\be {\cal L}=\ln \left((t_B-t_A)^2-(x_A-x_B)^2\right)\,,\ee
which gives the correct answer for the two-point correlation function of $(1+1)$-dimensional $CFT$:\footnote{If $\Delta x^2>\Delta t^2$, the renormalized length
is \be {\cal L}=\ln \left(-(t_B-t_A)^2+(x_A-x_B)^2\right)\,,\ee
and the full Green function is
\be G(t_A,x_A;t_B,x_B)=\frac{1}{|(t_B-t_A)^2-(x_A-x_B)^2|^{\Delta}} \,.\ee}
\be G(t_A,x_A;t_B,x_B)=e^{-\Delta {\cal L}}=\frac{1}{\left((t_B-t_A)^2-(x_A-x_B)^2\right)^{\Delta}}\,,\,\,\,\,\,\,\Delta t^2>\Delta x^2\,.
\label{eq:PatchGreen}\ee

In the global $AdS$ space-time the Poincar\'e horizon has no special physical meaning, but we can still try to generalize this procedure to
this case.

The boundary field theory now is defined on $S^1\times \mathbb{R}^1$ spacetime, and the two-point Green function
that we must be able to reproduce via the geodesic approximation has the form \cite{DiFrancesco}:
\be G(t_A,\phi_A;t_B,\phi_B)=\frac{1}{|\cos (t_B-t_A)-\cos (\phi_B-\phi_A)|^\Delta}\,. \label{GlobalGreen}\ee
Note that this function is periodic both in angle and time. While the angular periodicity is obvious by construction,
periodicity in time emerges because of the finite size effects: an excitation created at some point in space and time starts dissipating, but
later recollects and revives due to the spatial periodicity.

We will need a function that defines angular separation between boundary points while properly maintaining
the rotational invariance
of the system. For instance, given two angular coordinates $\phi_2=\frac{7\pi}{4}$ and $\phi_1=\frac{\pi}{4}$, the difference
between them along the shorter arc is \be D (\phi_2,\phi_1)=-\frac{\pi}{2} \neq \phi_2-\phi_1\,. \ee
Thus we should use
\begin{eqnarray}
 D(\phi_1,\phi_2)=\mod (\phi_2-\phi_1+\pi,2\pi)-\pi\,.\label{eq:difference}
\end{eqnarray}
Analogously, for the arithmetic average of two angular coordinates (that provides a point exactly at the middle of the shorter arc between $\phi_1$ and $\phi_2$):
\begin{equation}
 \Sigma(\phi_1,\phi_2)=
 \frac12 \left(\phi_1 + \phi_2 - 2 \pi \theta\left(-\cos\left(\frac12 \left(\phi_1 - \phi_2\right)\right)\right)\right)
\end{equation}

Note that the Green function \eqref{GlobalGreen} has a symmetry:
\be G(t_A,\phi_A;t_B,\phi_B)=G(t_A,\phi_A;t_B+\pi,\phi_B+\pi)\,,\ee
where points on the r.h.s. can be spacelike separated while points on the l.h.s.
have timelike separation:
\be (t_B-t_A)^2 > D (\phi_B,\phi_A)^2\,,\,\,\,\mbox{but} \,\,\,(t_B-t_A+\pi)^2 < D (\phi_B+\pi,\phi_A)^2 \,.\ee

This symmetry can be used to construct a disjointed spacelike geodesic, with two branches reattached at the Poincar\'e horizon,
connecting timelike separated points.

\begin{figure}
\begin{center}
\includegraphics[width=0.5\textwidth]{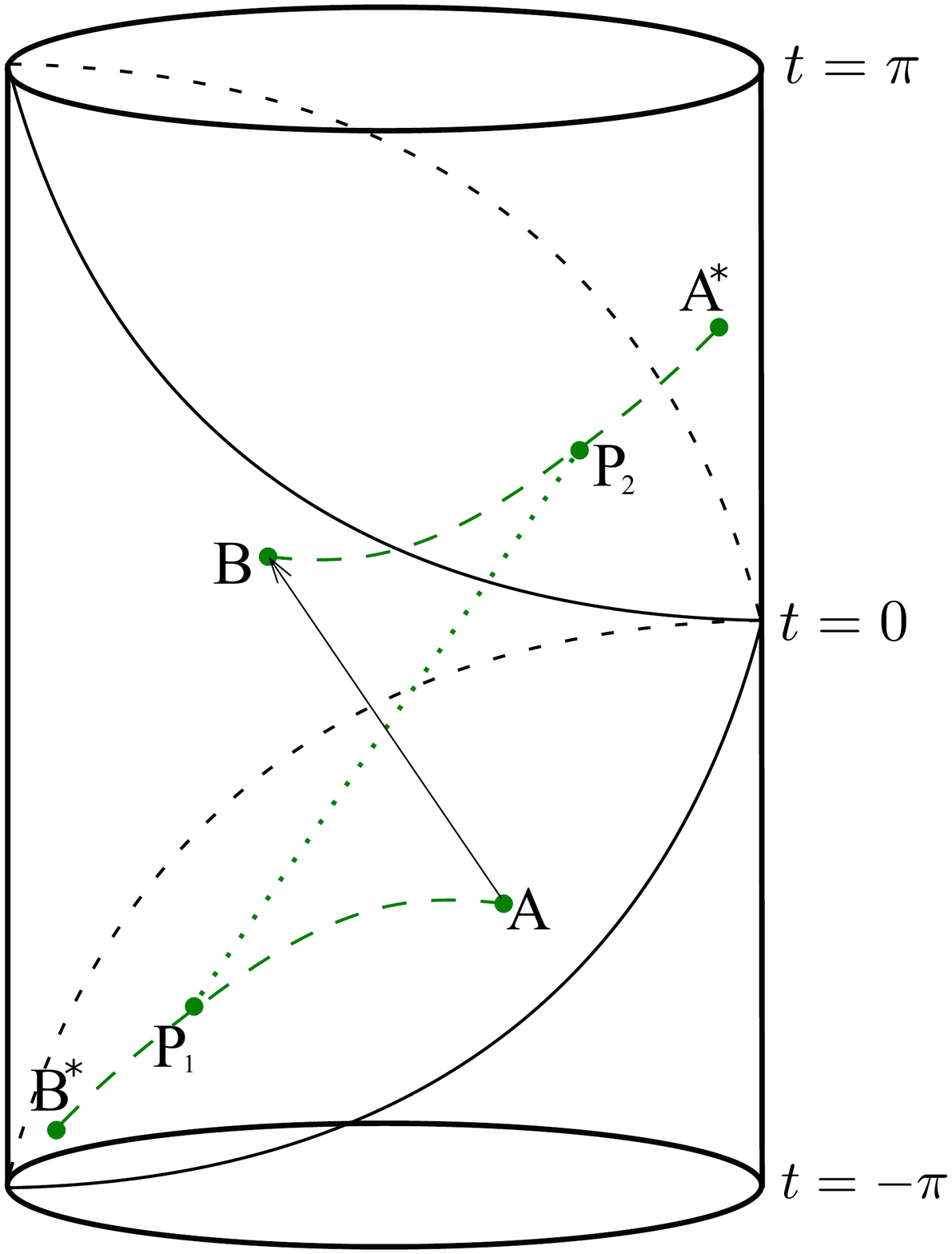}
\caption{A spacelike quasigeodesic connecting two boundary timelike separated points $A$ and $B$. A spacelike curve emerges from point $A$ and reach the Poincar\'e
horizon at point $P_1$. Then it jumps to a mirror point $P_2=P_1+(\pi_t,\pi_\phi,0_R)$ on the second plane of the horizon, and proceeds further to $B$.
The arc $P_2B$ is a rotation of $P_1 B^*$ by $\pi$.}
\label{fig:HorizonJump}
\end{center}
\end{figure}

If we represent the global $AdS_3$ space-time as a cylinder, the Poincar\'e horizon consists of two planes cutting the cylinder
at $45^\circ$. The orientation of the planes (as a rigid construction) can be chosen arbitrarily.
Then consider a spacelike geodesic emerging from boundary point $A$ (see Fig.\ref{fig:HorizonJump}), and terminating at boundary point $B^*$.
Somewhere in the bulk it has a turning point $P_1$ where its radial coordinate $R^*=R(0)$ is minimal.

Since we are free to choose the location of the Poincar\'e horizon, we can always orient it in such a way that the $AB^*$ geodesic
intersects it at the turning point $P_1$. This point $P_1$ can be identified with a point $P_2=P_1+(\pi_t,\pi_\phi,0_R)$, located on the other
cutting plane. The arc $P_1B^*$ can be then rotationally translated to this point: $P_1\rightarrow P_2$, and then $B^* \rightarrow B$, where
$B=B^*+(\pi,\pi)$.
Hereafter we will use disjointed ``quasigeodesics'' of this $AP_1P_2B$ type to connect timelike separated points.

The explicit analytic expression for the quasigeodesics can be derived in the following way.
Let's take the Poincar\'e chart geodesics \eqref{eq:Belgian}, and rewrite them in coordinates of the embedding spacetime.
The corresponding coordinate transformation is given by:
\bea X^0&=& \frac{r}{2}(\frac{1}{r^2}+1+x^2-t^2),\\
X^1&=&xr,\\
X^2&=&\frac{r}{2}(\frac{1}{r^2}-1+x^2-t^2),\\
X^3&=& rt\,.\eea
Substituting here \eqref{eq:Belgian}, and recalling the relations between the global and embedding coordinates \eqref{eq:SchwarzschildGlobal}, we obtain
\bea \sqrt{1+R^2}\cos t  &=& \frac{(-1+E^{2}-J^{2})}{2\sqrt{E^{2}-J^{2}}}\sinh\lambda\,,\\
\sqrt{1+R^2}\sin t  &=& \frac{E}{\sqrt{E^2-J^2}}\cosh\lambda\,, \\
R\cos\phi &=&\frac{J}{\sqrt{E^{2}-J^{2}}}\cosh\lambda\,, \\
R\sin\phi &=&\frac{(-1-E^{2}+J^{2})}{2\sqrt{E^{2}-J^{2}}}\sinh\lambda\,.  \eea
These can be solved to give us the embedding of the Poincar\'e chart spacelike geodesic into the global $AdS$:
\bea t (\lambda)&=&\arctan\left( \frac{2E}{-1+E^2-J^2}\coth\lambda\right)+t _0,\\ \label{eq:quasigeodesic}
 \phi(\lambda)&=&\arctan\left( \frac{-1-E^2+J^2}{2J}\tanh\lambda\right)+\phi_0,\\
 R(\lambda)&=&\sqrt{\frac{J^2}{E^2-J^2}\cosh^2\lambda+\frac{(-1-E^2+J^2)^2}{4(E^2-J^2)}\sinh^2\lambda}\,. \label{eq:Rlambda}\eea
This quasigeodesic already has a $\pi$-jump in time at the turning point $\lambda=0$, and as explained before we also need to adjust the discontinuity in angle:
\be \widetilde{\phi}(\lambda)=\phi(\lambda)+\pi\theta(\lambda)=\arctan\left(\frac{-1-E^2+J^2}{2J}\tanh\lambda\right)+\pi\theta(\lambda)+\phi_0\,.\ee
From now on we will omit the tilde.

The kinetic invariants can then be expressed in terms of the boundary coordinates:
\bea E &=&\frac{\sin \frac{t_2-t_1}{2}}{\sin \frac{D(\phi_2-\pi,\phi_1)}{2}-\cos \frac{t_2-t_1}{2}}\,, \label{eq:Kinetics} \\
J &=& \frac{\cos \frac{D(\phi_2-\pi,\phi_1)}{2} }{\sin \frac{D(\phi_2-\pi,\phi_1)}{2} -\cos \frac{t_2-t_1}{2}}\,.
\eea
The integration constants $t _0$ and $\phi_0$ in \eqref{eq:quasigeodesic} can be represented as:
\bea
t_0=\frac{1}{2}\left(t_1+t_2\right)\,,
\phi_0=\Sigma(\phi_1,\phi_2-\pi)\,.
\eea

Inverting equation \eqref{eq:Rlambda} we obtain dependence of the affine parameter on the radial coordinate:
\be \lambda (R) =\pm\arcsinh\sqrt{\frac{4(E^2-J^2)R^2-4J^2}{(-1-E^2+J^2)^2+4J^2}}\,,\ee
where the minus sign is taken on the first branch of the geodesic or quasigeodesic (i.e. before the turning point,
- when particle moves away from the boundary), and the plus sign is taken on the second branch (when particle moves towards the boundary).
This function can be used to define the geodesic length, which is simply
\be \cL (R_1,R_2)=\lambda_\pm(R_2)-\lambda_\pm(R_1)\,,\ee
for two points with radial coordinates $R_1$ and $R_2$.

Note that for a geodesic connecting two boundary points the length is divergent:
\be \cL=\lim\limits_{R\rightarrow\infty}(\lambda_+(R)-\lambda_-(R))=2\lim\limits_{R\rightarrow\infty}\sqrt{\frac{4(E^2-J^2)R^2-4J^2}{(-1-E^2+J^2)^2+4J^2}} =\infty\,,\ee
and needs to be renormalized. The natural way to do it is to subtract the parameter independent divergent part, and define the geodesic length as
\be {\cal L}_{ren}=\lim\limits_{R\rightarrow \infty}\left(\lambda_+(R)-\lambda_-(R)-2\ln R\right)=\ln\left(\frac{16(E^2-J^2)}{E^4-2E^2(-1+J^2)+(1+J^2)^2}\right)\,. \label{eq:RenormGeodesic}\ee
The argument of the logarithm is always positive for quasigeodesics connecting timelike separated points, but can be less than $1$. It means that the geodesic length after the renormalziation in principle can be negative.
Using \eqref{eq:RenormGeodesic} along with \eqref{eq:Kinetics} we obtain the correct result for the retarded Green function:
\bea G_c(t_1,\phi_1; t_2,\phi_2)&=&e^{-\Delta \cL_{ren}}=\frac{1}{(\cos(t_2-t_1)-\cos(\phi_2-\phi_1))^\Delta}\,,\label{eq:CylindricGreen}\\
\mbox{where}& & \,\,\, (t_2-t_1)^2>D(\phi_1,\phi_2)^2\,,\nonumber \eea
where $D(\phi_1,\phi_2)$ is the function introduced in \eqref{eq:difference}.
The possible negativity of the renormalized geodesic length is the reason why singularities of the correlator can be
captured in the geodesic approximation.

Here we must pause for a second and stress Lorentz non-invariance of \eqref{eq:CylindricGreen}.
We define $SO(2,2)$ isometries of $AdS_3$ in terms of the Lornetz boosts of the embedding
$\mathbb{R}^{(2,2)}$ space \eqref{eq:EmbedLorentz}. If we take two boundary points $A$ and $B$, and act on them with a bulk isometry transformation $\Lambda$
of this type, we will observe that it does not preserve the Green's function \eqref{eq:CylindricGreen}:
\be G_c(\Lambda A, \Lambda B)\neq G_c(A,B)\,.\ee
It is the fundamental difference between holography of a Poincar\'e chart and holography of global $AdS$. In the first case the bulk isometries
induce Lorentzian boosts on the boundary, so the Green function of a
dual boundary field theory is a relativistic invariant object \eqref{eq:PatchGreen}.
In the second case the isometries rather act as conformal transformations leaving the Green function {\it covariant}, i.e. invariant up to some coordinate dependent scaling prefactors.

In the holographic language this is encoded in the fact that the renormalized lengths connecting boundary points are dependent on the choice of the reference frame.
Below when we consider the DeDeo-Gott time machine geometry, we should be especially careful about this, since the geodesics there are combinations of Lorentz invariant and non-invariant terms as, for example, in \eqref{eq:SumOfArcs}.
The proper way to deal with it is explained in the first subsection of Sec. \ref{sec:Green}.

In the next section we will analyze lensing of the quasigeodesics on the conical defects and calculate the Green function of the dual field theory
in presence of the closed timelike curves in the bulk.

\section{The two point Green's function \label{sec:Green}}

\subsection{The algorithm}

In Sec. \ref{sec:Entwinement} we have discussed the general idea of using the geodesic approximation to compute the boundary Green's
function for the DeDeo-Gott geometry. Now we will formulate an exact algorithm for that.

\begin{itemize}
 \item Introduce coordinate system on the unfolding of the double-cone space in such a way, that the physical (unremoved)
 part of the boundary consists of two stripes covered by coordinate intervals:
 \bea t &\in& \left(-\infty,\infty \right)\,,\\
 \phi &\in& \left(-\frac{\pi}{2}-\frac{\Delta\phi}{2}+t,-\frac{\pi}{2}+\frac{\Delta\phi}{2}+t\right)
 \cup \left(\frac{\pi}{2}-\frac{\Delta\phi}{2}+t,\frac{\pi}{2}+\frac{\Delta\phi}{2}+t\right)\,. \nonumber \eea
 \item Fix two boundary points $A$ and $B$. For simplicity we can choose $A=(0,-\frac{\pi}{2})$.
 \item Fix the total number of windings $N$ that a geodesic of interest undergoes on the way from $A$ to $B$.
 In our simulations we will not go beyond $N=4$, because the higher-order contributions to the Green's function are highly suppressed.
 \item For the given $A$, $B$, and the number $N$, generate all possible $4\cdot 3^{N-1}$ sets of the complementary points $\left\{C_1,\cdots C_{2N}\right\}$
 corresponding to different sequences of windings $\left\{ \cW_1,\cdots \cW_N\right\}$. The (quasi)geodesics then consist of $N+1$ arcs $AC_1$, $C_2C_3$,
 ..., $C_{2N}B$, each of which is just a (quasi)geodesic curve in empty $AdS_3$.
 \item Impose that each of the ``odd'' complementary points $C_{2i+1}$ belongs to the causal future of the previous
 ``even point'': $C_{2i+1} \succ C_{2i}$.

 Let us elaborate on what the reason to do so is. The most clear question we can ask is whether {\it causal} propagation of a signal from the future to the past is possible.  To define the dual retarded Green function in presence of the CTC in the bulk, we should recall that evolution of a particle moving in the bulk of $AdS$  can be split in two parts: ``physical'' continuous motion along a timelike or a spacelike geodesic, and ``topological'' time jumps caused by entwinement  around the conical defects. In the holographic language geodesic branches $C_{2i}C_{2i+1}$ correspond to the continuous evolution, and
 $C_{2i+1}\rightarrow C_{2i+2}$ identifications - to the time jumps. In absence of the closed timelike curves a signal could
 causally propagate from $A$ to $B$ if $B$ belongs to the future light cone of $A$: $B \succ A$. A natural generalization of this prescription
 for the time machine case is to impose that this should hold true for all ``physical'' segments, i.e. $C_{2i+1} \succ C_{2i}\,\,\,\forall\,\, i$.
  \item For each of the causal quasigeodesics, solve for the intersection points $\left\{ O_1, \cdots O_{2N}\right\}$.
 The easiest way to do this is to transform for each winding back to the rest frame of the corresponding wedge.
 For example, if branch $C_{2i}C_{2i+1}$ intersects first the trailing face of the 2nd wedge, and then the leading face of the 1st wedge,
 we perform a Lorentz transformation of the branch to the 2nd rest frame, then untwist the wedge by a simple coordinate transformation
 $\phi^\prime=\phi-t$, such that angular location of the face remains still in these co-rotating coordinates, and solve the equation
 \be (\phi_{C_{2i}C_{2i+1}}^{II}(\lambda_{2i})-t_{C_{2i}C_{2i+1}}^{II}(\lambda_{2i})) \mod 2\pi=\phi^\prime_{T_{II}} \mod 2\pi \,.\ee
 Then we repeat the procedure in the 1-st rest frame:
 \be (\phi_{C_{2i}C_{2i+1}}^{I}(\lambda_{2i+1})-t_{C_{2i}C_{2i+1}}^{I}(\lambda_{2i+1}))\mod 2\pi=\phi^\prime_{L_{I}} \mod 2\pi \,.\ee
 \item Make sure that all these equations have real solutions (otherwise discard the geodesic).
 \item Make sure that if a branch is not expected to intersect other faces within the physical region of the space, it actually does not
 (fake intersections within the removed part of the unfolding are allowed). In other words, if an arc $O_{2i}O_{2i}$ emerges from the face $L_{I}$
 and terminates at the face $T_{II}$, it should not have intersections with $L_{II}$ and $T_I$.
 \item Calculate the lengths of all inner segments of the geodesic ($O_{2i}O_{2i+1}$). They are finite by construction and equal to
 \be L_{O_{2i}O_{2i+1}}=\lambda_{2i+1}-\lambda_{2i}\,.\ee
 \item Renormalize the lengths of the boundary segments $AO_1$ and $O_{2N}B$ as they are divergent:
 \bea L_{AO_1}=\frac12 L^{ren}_{AC_1}+\lambda_1, \\
 L_{O_{2N}B}=\frac12 L^{ren}_{C_{2N}B}-\lambda_{2N}\,.
 \eea
 \item  Calculate the renormalized lengths of $L_{AO_1}$ and $L_{O_{2N}B}$ in the {\it original frame}. As mentioned in the previous section, the renormalized lengths are not Lorentz-invariant.
 So, while we are free to constantly switch between different reference frames in order to calculate lengths of the finite inner segments
 $O_{2i}O_{2i+1}$, the renormalized lengths of the two boundary segments must be calculated in the original frame where we define the Green's function.
 In our case it is the ``centre-of-mass frame'', where the two conical defects are symmetrically boosted.
  \item Finally calculate contribution of the geodesics to the Green function:
 \be G(A,B)=\sum\limits_{k}e^{-\Delta L_k}\,,\ee
 where the index $k$ runs over the set of geodesics that satisfy aforementioned conditions.
\end{itemize}

\begin{figure}
\begin{center}
\begin{tabular}{cc}
\includegraphics[width=0.5\textwidth]{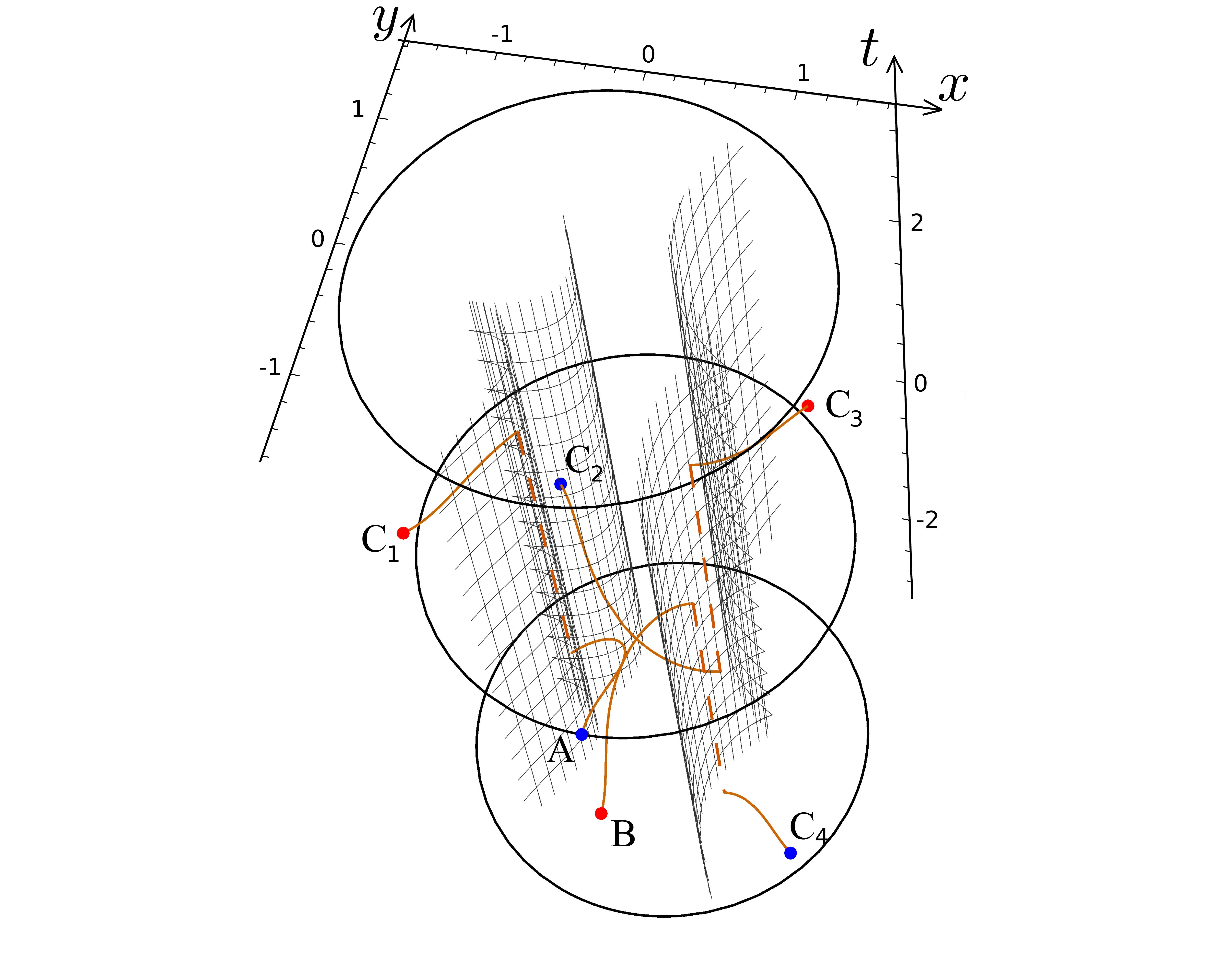}
\includegraphics[width=0.5\textwidth]{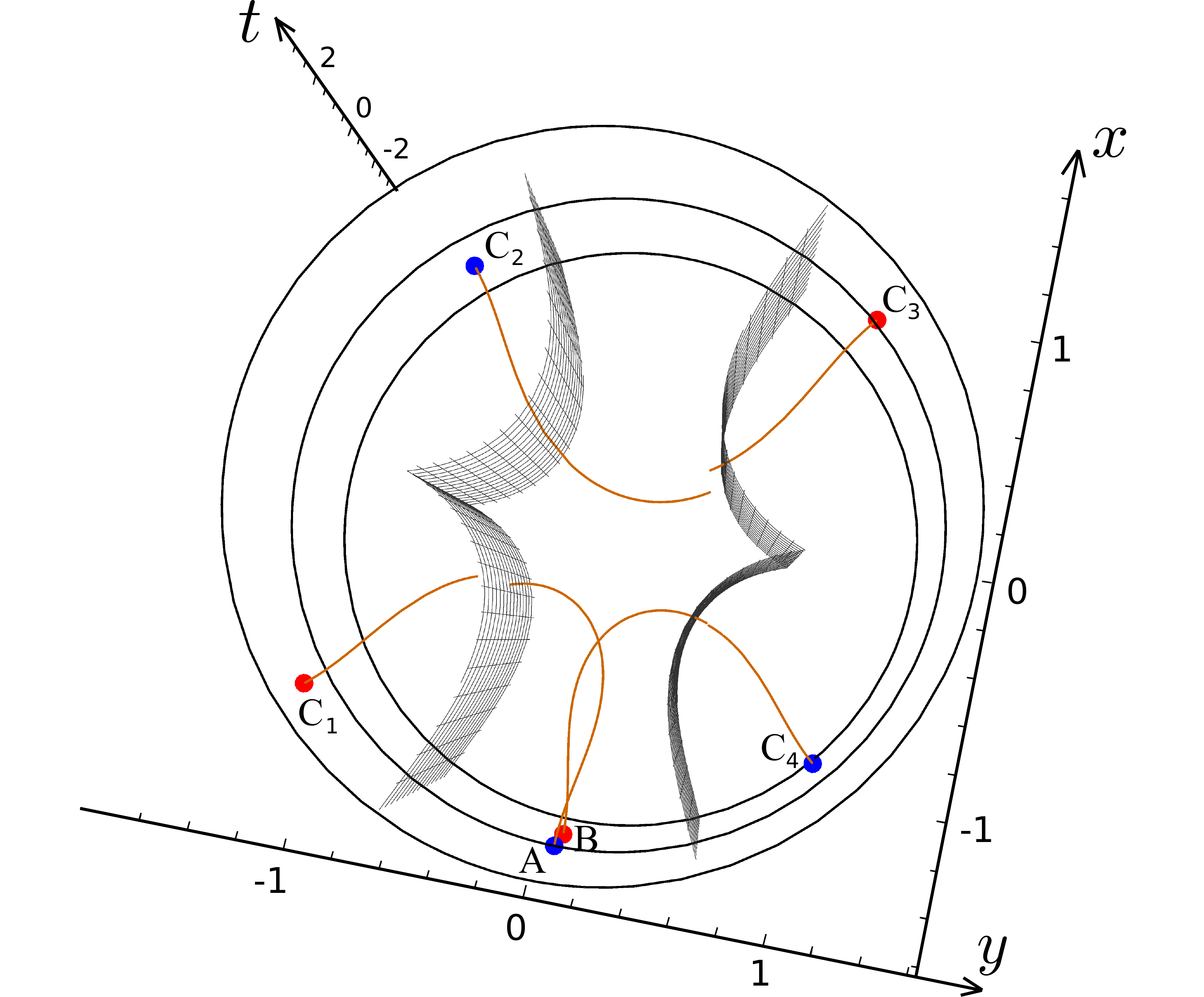}
\end{tabular}
\caption{A quasigeodesic with winding number $N=2$ connecting boundary points $A$ and $B$.}
\label{fig:N2winding}
\end{center}
\end{figure}
In accordance with the described algorithm we subsequently account for
contributions from higher winding numbers starting with $N=2$\footnote{For negative times $N=1$ windings do not contribute as
they are due to the lensing on a single conical defect, that obviously can not lead to time travelling. But for positive times we take them
into account.}.
In other words, we formulate a kind of ``perturbation theory'' with the number of entwinements as a control parameter.

One property of this series expansion must be comment on. Each geodesic contributes to the Green function exponentially:
\be e^{-\Delta L_{ren}}\,.\ee
For higher windings the number of internal segments $C_{2i}C_{2i+1}$ of the geodesic grows linearly in $N$, and so does
its renormalized length $L_{ren}$. Therefore the corresponding contribution to the Green function is exponentially small\footnote{In a generic case
when $L_{ren} >0$.}.
On the other hand the total number of possible topologically different geodesics scales
as
\be n \sim 3^{N-1} \,,\ee
i.e. grows exponentially. Therefore in principle these two effects can compete and we can not say a priori that
the higher order contributions to the Green function are suppressed, and the sum over entwinements is convergent.
If not, this could mean that our setup is unstable and undergoes a Hagedorn like transition.
\begin{figure}
\begin{center}
\begin{tabular}{cc}
\subfigure[]{\includegraphics[width=0.5\textwidth]{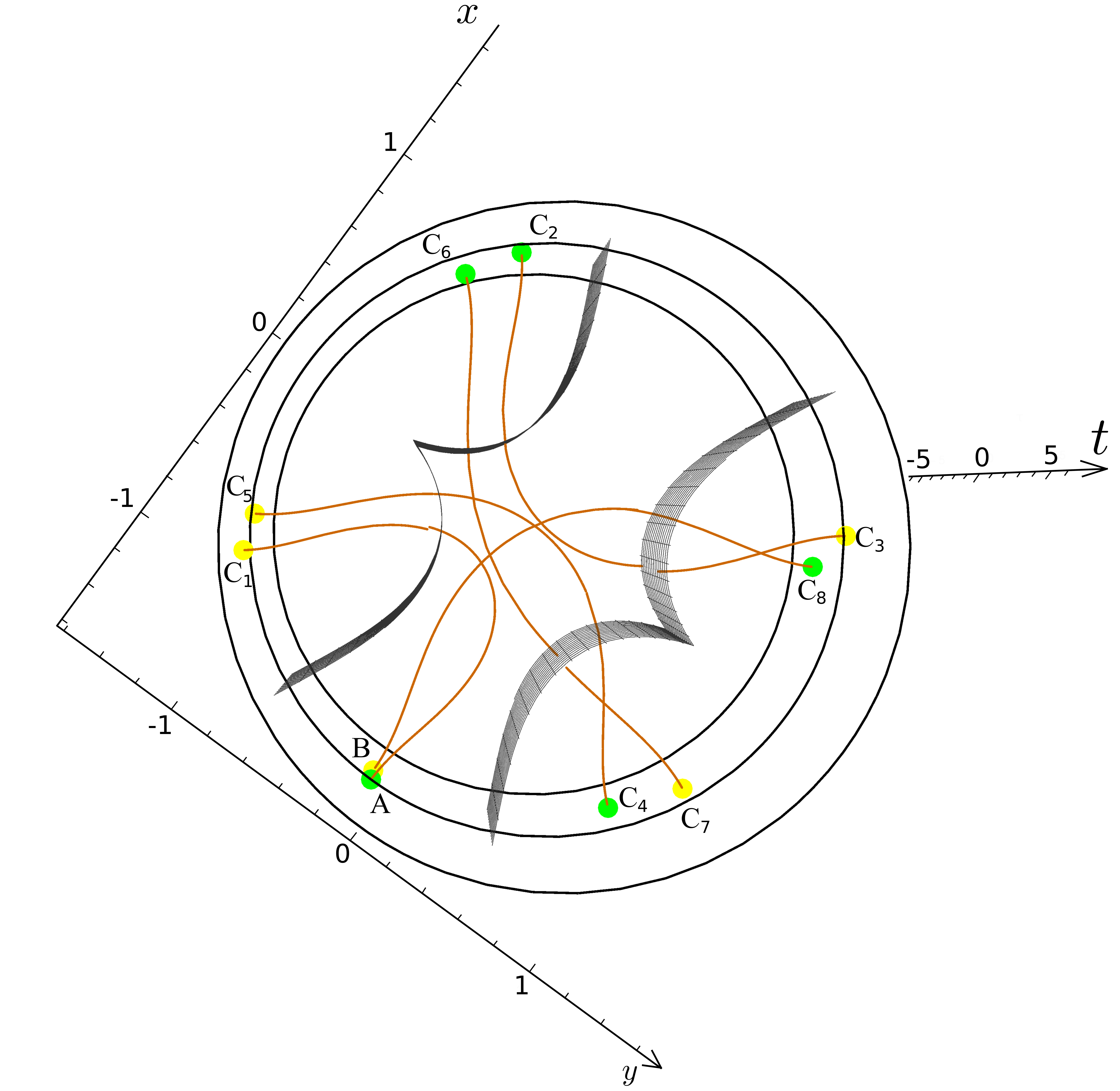}}
\subfigure[]{\includegraphics[width=0.5\textwidth]{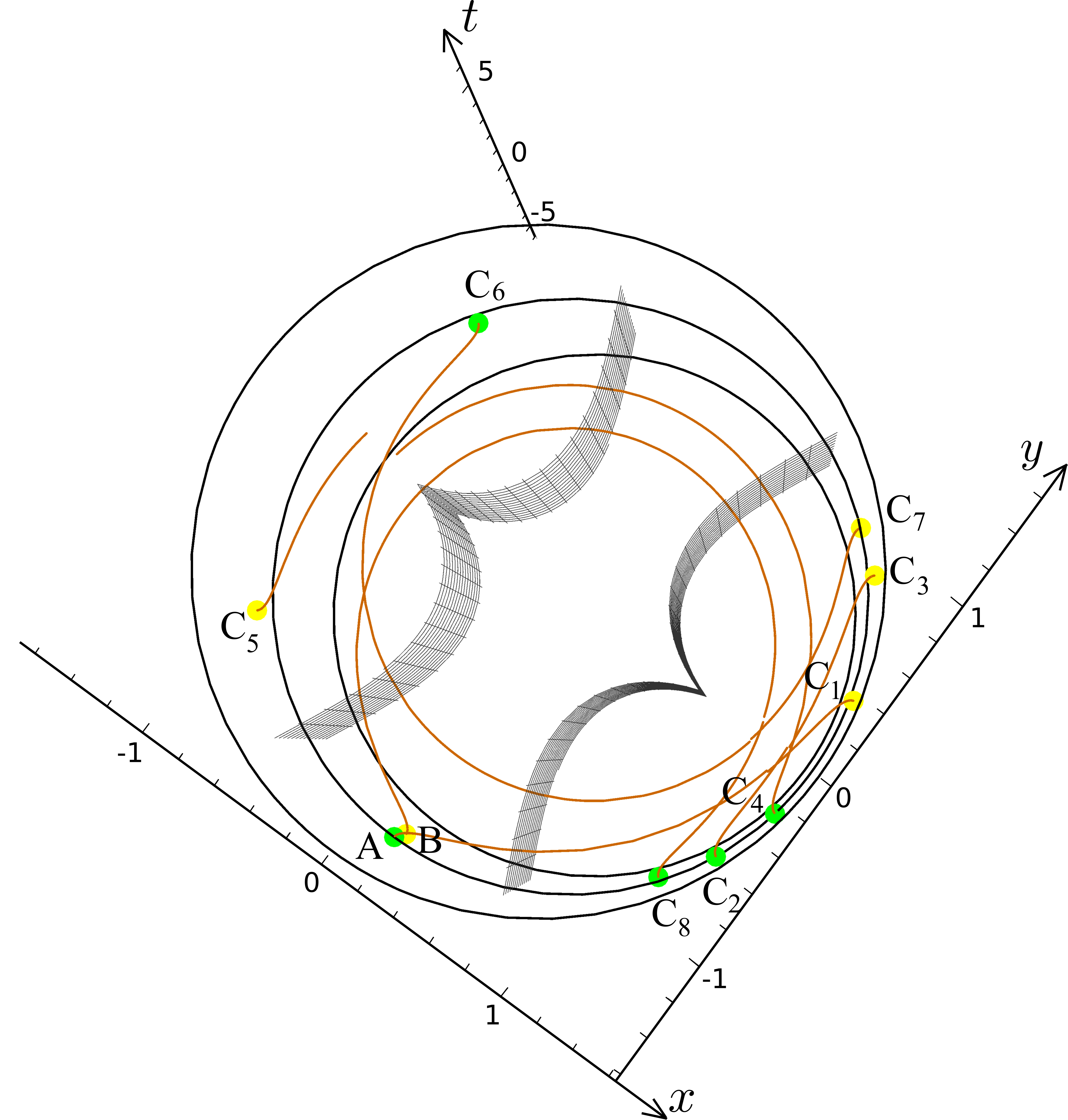}}
\end{tabular}
\caption{(a) An example of a physical $N=4$ geodesic contributing to the Green function. (b) An example of physically impossible winding:
for a given sequence of entwinements a geodesic can not be fit in the unremoved part of the spacetime.}
\label{fig:PhysicalUnphysical}
\end{center}
\end{figure}

However there are three different reasons for it not to happen. Firstly, by no means all of these $4\cdot 3^{N-1}$ winding configurations
satisfy the causality condition: $C_{2i+1} \succ C_{2i}$. Secondly, even if the causality condition for the set of complementary points is satisfied,
the geometric structure of the geodesics becomes more and more complicated as the number of windings increases, and it becomes hard to force a geodesic
curve to undergo the concrete sequence of windings (it is easy to see on Fig.\ref{fig:PhysicalUnphysical}(b)). Finally, the ``decaying'' exponent has a
conformal dimension as a knob, so at large enough $\Delta$ it dominates over the ``growing'' exponent.

Another way to understand convergence of the series expansion in all orders relies on a simple and general argument.
Consider a germ of all possible quasigeodesics emerging from point $A$. The first segment of a generic quasigeodesic
curve hits the boundary at some point $C_1$, first in the sequence of complementary points $\left\{C_i\right\}$. If we go along the curve further, we will
obviously see that it is defined uniquely up to the final moment when it reaches the physical part of the boundary at point $B$. Thus, for a fixed initial
point $A$, for each of the ``first-in-the-sequence'' complementary points $C_1$ the final point $B$ is defined unambiguously.
Now, as we have already emphasized, the singular contributions to the Green function come at the points where the renormalized geodesic length
is infinitely negative, $\cL^{ren}=-\infty$. It is possible if and only if $C_1$ is located exactly on a generatrix of the light cone
emerging from point $A$, or $C_2N$ is located on the generatrix of the light cone of point $B$. Quasigeodesics having the complementary points $C_1$ and
$C_{2N}$ right on the corresponding light cones {\it form a zero measure subset} among all possible quasigeodesics. Thus the set of boundary points where the Green function is infinite is also a zero measure subset of the boundary spacetime. Everywhere else the Green function is finite and well-defined.


\subsection{Phenomenology and discussion}
We are now ready to implement our computational algorithm for the DeDeo-Gott geometry.
As shown in Sec. \ref{sec:TM}, closed timelike curves in the spacetime are present when
the total angular deficit is more than $2\pi$. For concreteness we impose $\alpha_{I,II}=\sqrt{3}\pi$,
and the boost rapidities $\psi_{I,II}=\pm1$. In their corresponding rest frames (in the co-rotating coordinates) the locations of the edges are taken to be
\be
\begin{matrix}
\phi^\prime_{L_1}&=& \alpha/2 & \phi^\prime_{T_1} &=& -\alpha/2,\\
\phi^\prime_{L_2}&=& \alpha/2 + \pi & \phi^\prime_{T_2} &=& -\alpha/2+\pi\,.
\end{matrix}
\ee
It is more convenient to calculate the Green function also in the co-rotating coordinates:
\be G^{cr}(t_1,\phi^\prime_1;t_2,\phi^\prime_2)=G(t_1,\phi_1-t_1;t_2,\phi_2-t_2)\,.\ee
\begin{figure}
\begin{center}
\includegraphics[width=0.5\textwidth]{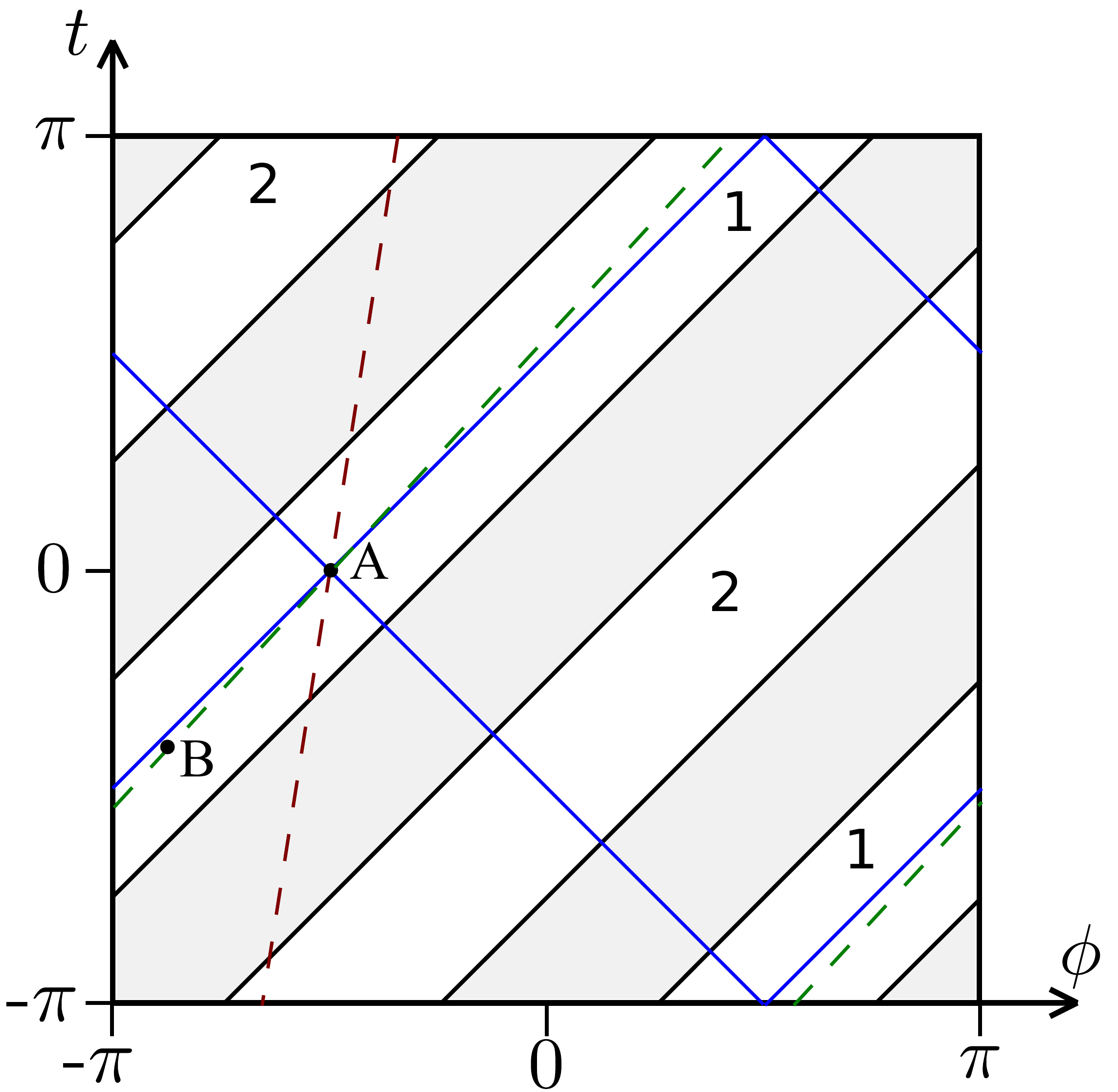}
\caption{The Green function is sourced at point $A$ with coordinates $(0,-\pi/2)$. Blue lines depict the light cone emerging from this point.
Any timelike line would cross the removed region and enter another strip (red dashed line). In order to avoid formulating the boundary field theory
on both strips simultaneously we calculate the Green function on a timelike line very close to the light cone generatrix.}
\label{fig:TimeLikeLine}
\end{center}
\end{figure}
For simplicity we will mostly study the Green function on a one-dimensional timelike line passing through the point $A$.
We should be careful here. Any timelike line originated in one physical strip crosses the cut out region and enters the second strip.
To formulate a quantum field theory on both strips simultaneously is possible yet tricky due to the fact that on the unification of two
parts of the boundary time can not be globally defined. To avoid this difficulty we will consider the Green function on a
timelike line in a close vicinity of the generatrix of the light cone. Then in a large range of times we will stay within one strip of the boundary.

In other words, the object we will attempt to evaluate is (in the co-rotating frame)
\be G^{cr}(0,-\pi/2;t,-\pi/2+\epsilon t)\,,\,\,\,\,\epsilon \ll 1\,.\ee

We have performed the numerical calculation of the retarded Green function for negative times $t<0$ up to $N=4$ order, and
for positive times $t>0$ up to $N=2$.

\begin{figure}[t]
\begin{center}
\begin{tabular}{cc}
\subfigure[]{\includegraphics[width=0.35\textwidth]{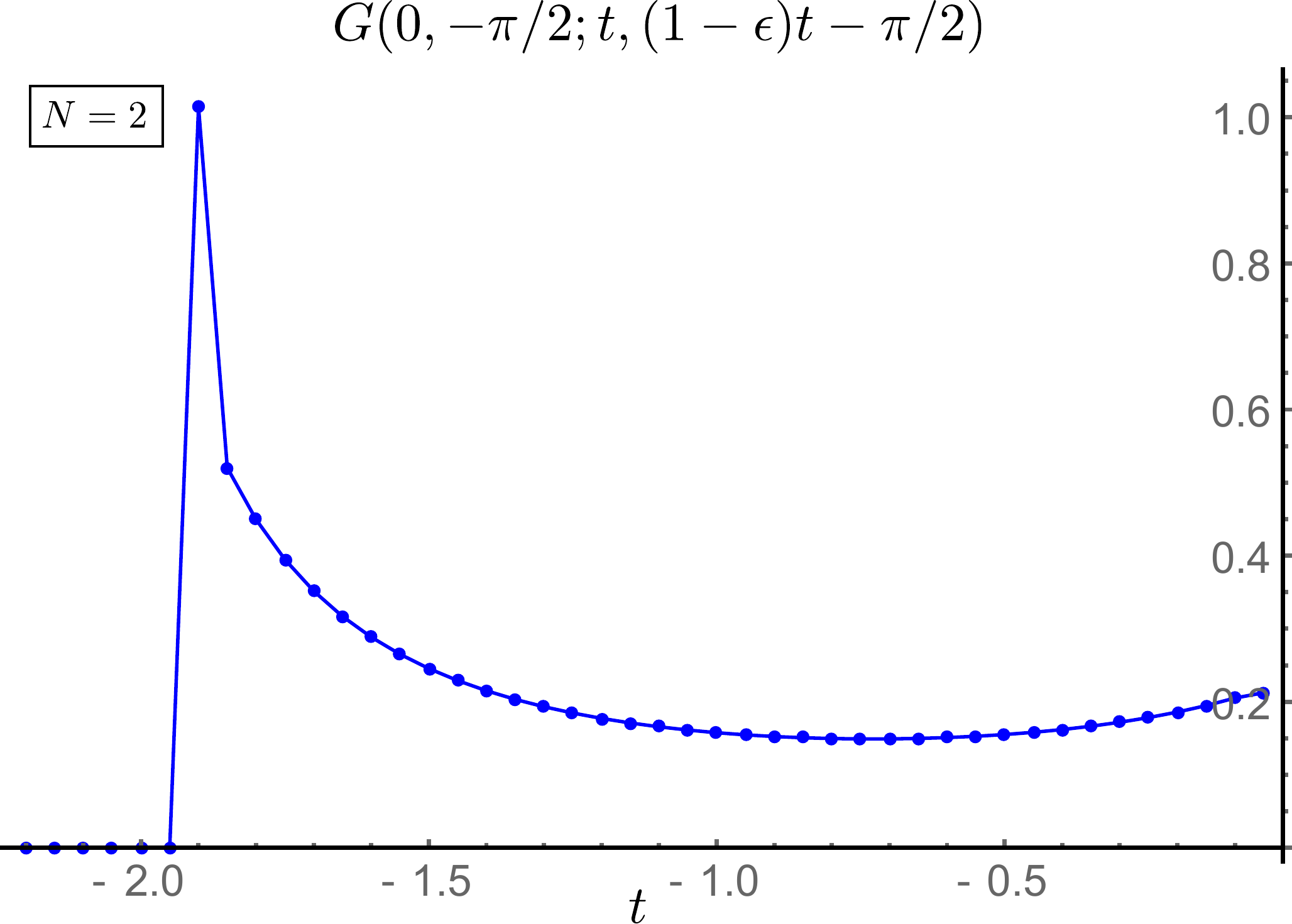}}
\subfigure[]{\includegraphics[width=0.35\textwidth]{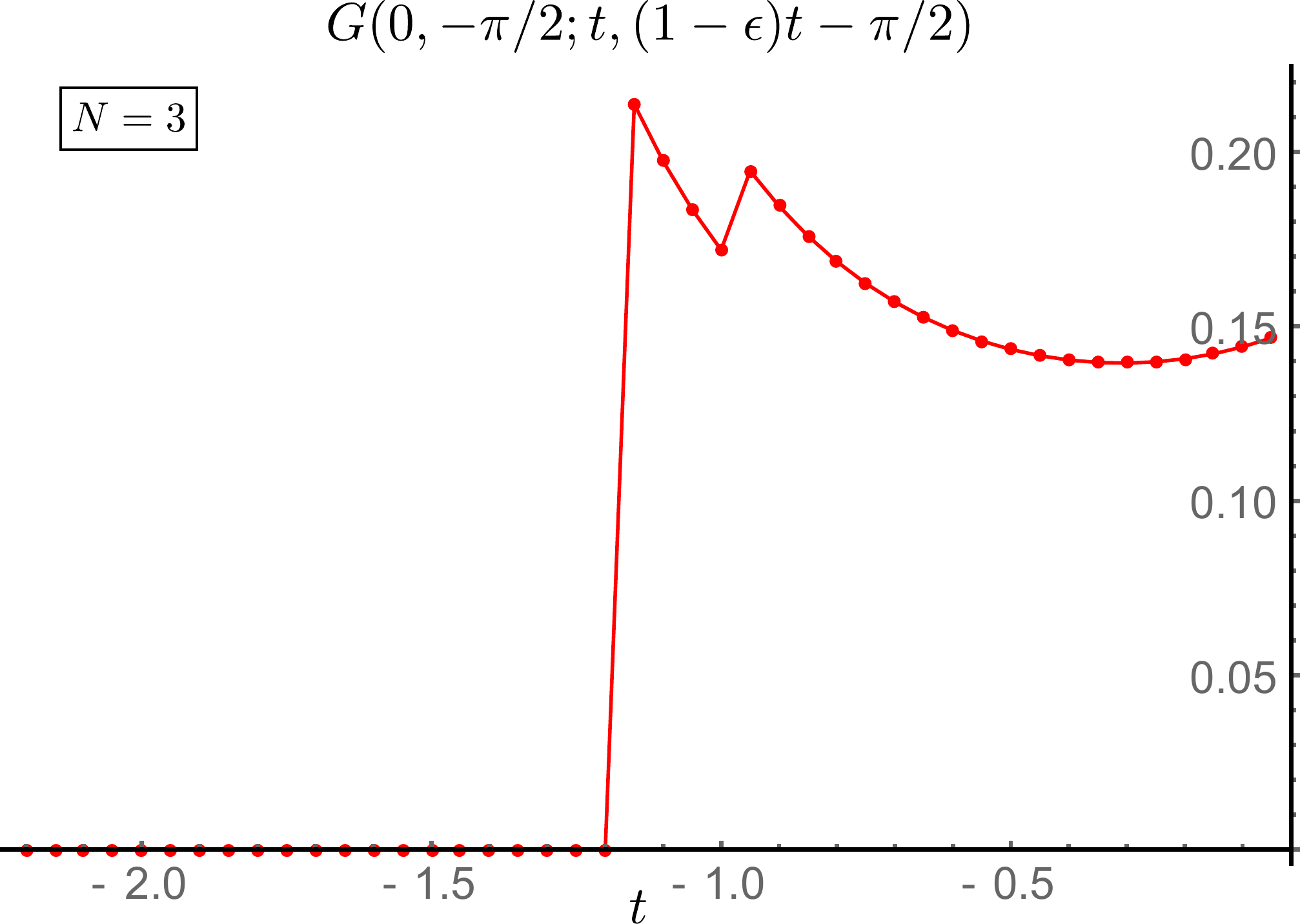}}
\subfigure[]{\includegraphics[width=0.35\textwidth]{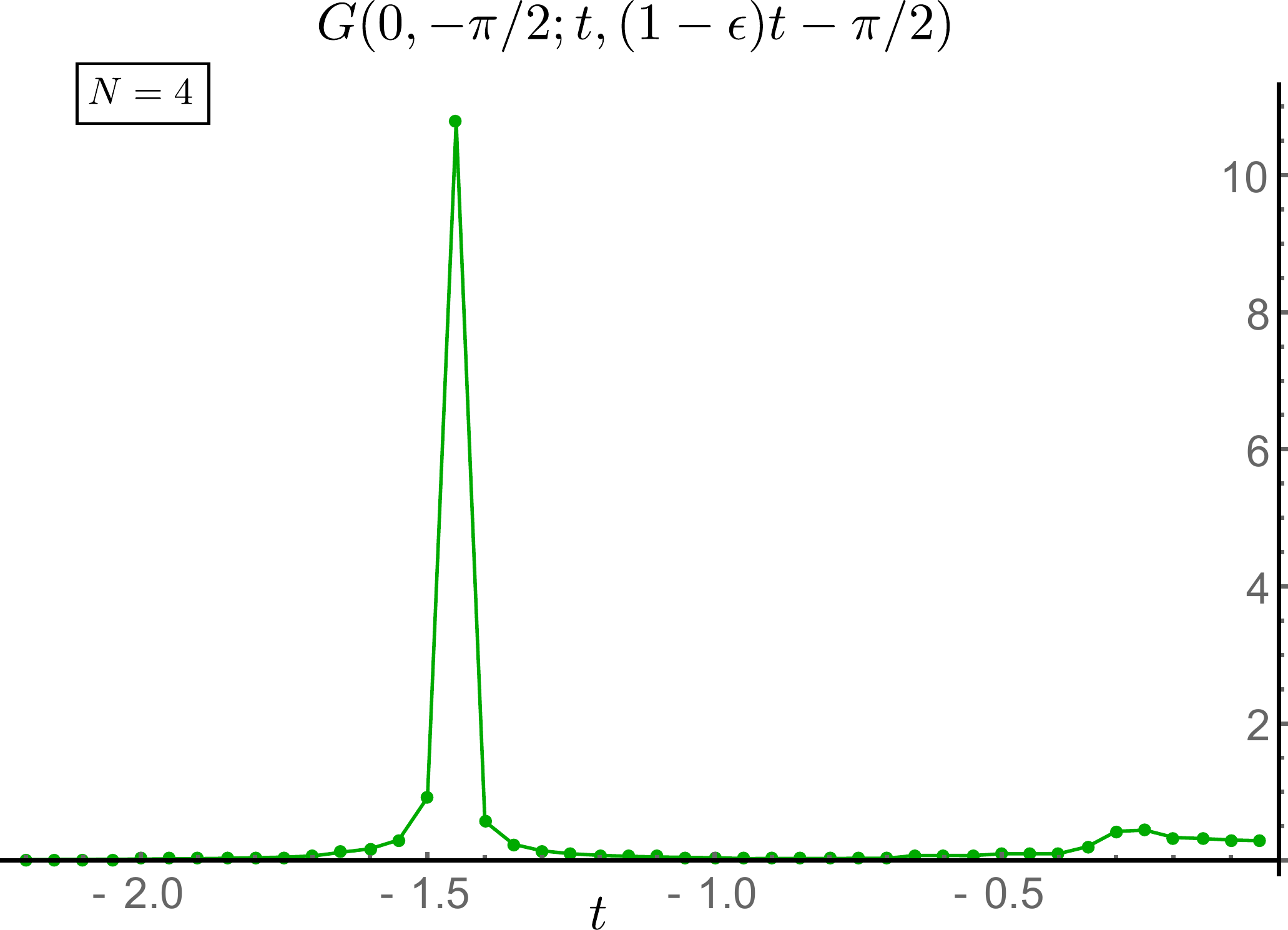}}\\
\subfigure[]{\includegraphics[width=0.7\textwidth]{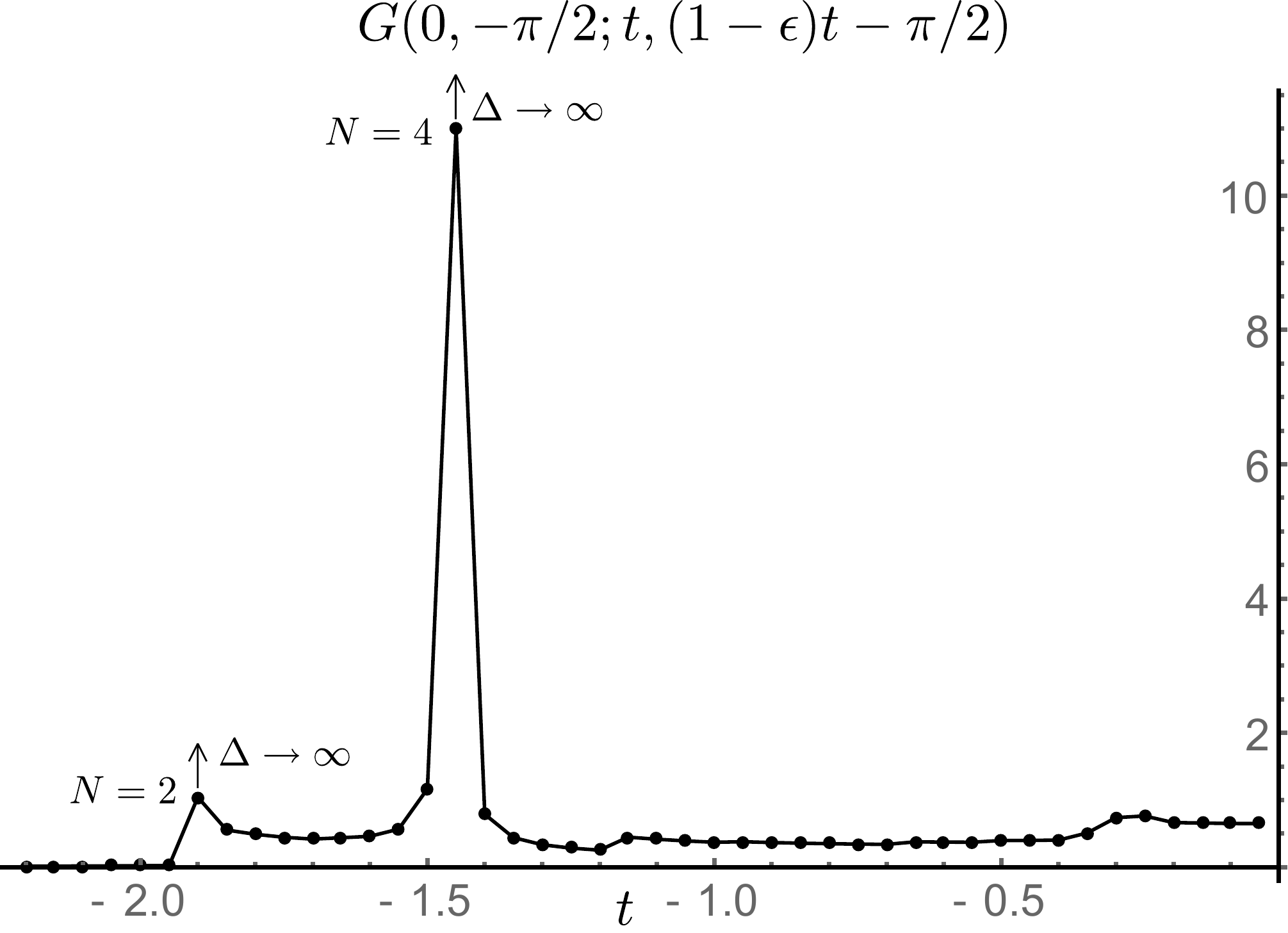}}
\end{tabular}
\caption{(a) $N=2$, (b) $N=3$ and (c) $N=4$ contributions to the retarded Green function at negative times at $\Delta=1.5$. Discontinuities of the curves are
artifacts of the geodesic approximation. (d)
The retarded Green function at $\Delta=1.5$ ($N=2$, $N=3$ and $N=4$ contributions are added up). For the large conformal dimensions peaks are enhanced, not suppresses,
and we can see revival of the particle at moments preceding the excitation of
the Green function. A not very large conformal dimension is chosen
for convenience of presentation. Here $\epsilon=0.1$.}
\label{fig:GreenFunctionDelta15}
\end{center}
\end{figure}
\begin{figure}[t]
\begin{center}
{\includegraphics[width=0.5\textwidth]{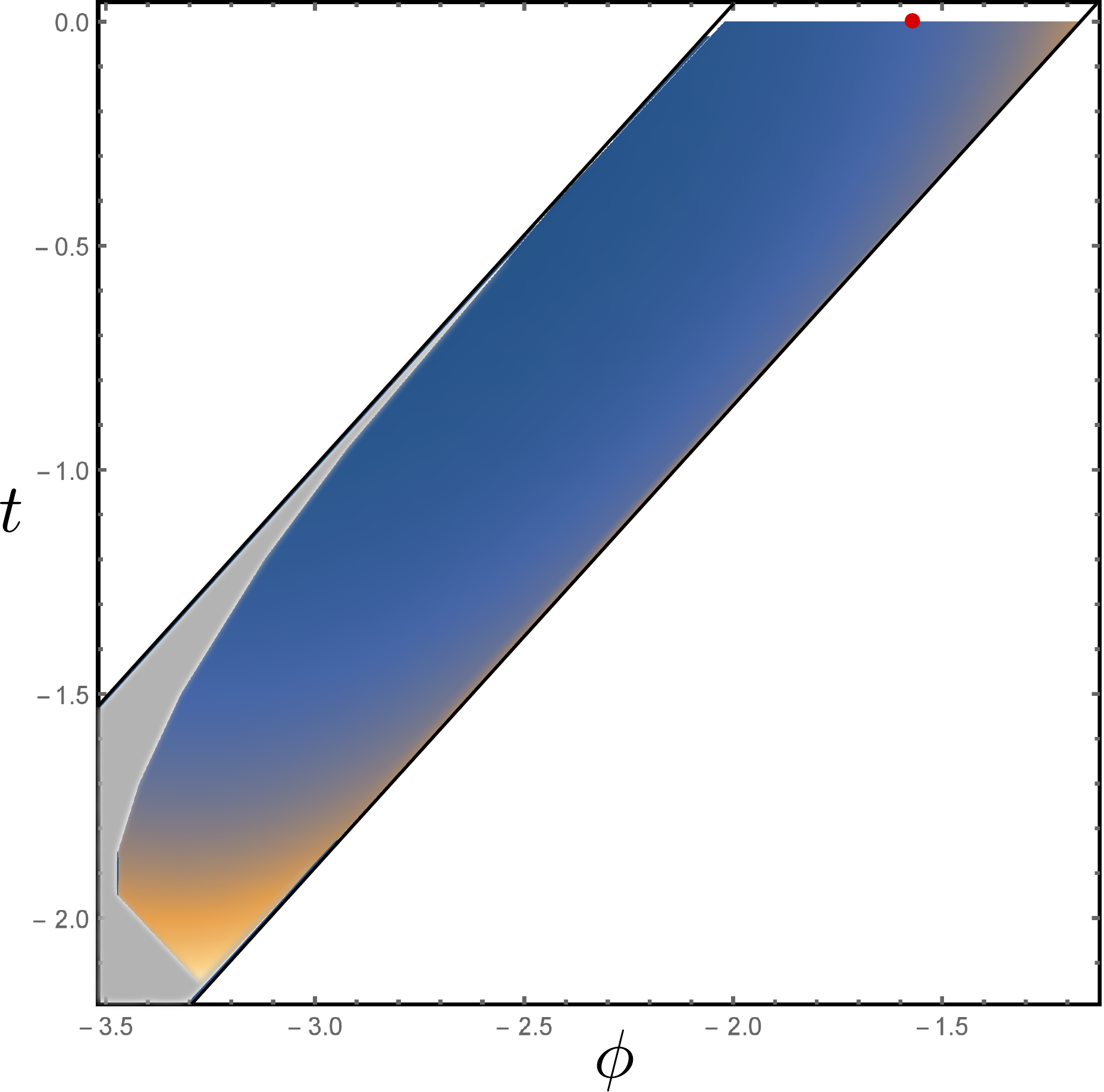}}
\caption{$N=2$ contribution to the retarded Green function at $\Delta=1.5$ at negative times in two dimensions.
The Green function is sourced at the red point $(0,-\pi/2)$. We construct the theory only within one of the two strips of the boundary. Sudden break of the function signalizes that some regions of the spacetime
are unattainable for the $N=2$ quasigeodesics.}
\label{fig:2DGreen}
\end{center}
\end{figure}

Let's discuss firstly the analytic behaviour of the Green function at negative times, - how the quantum particle behaves travelling back in time.
A naive expectation would be to think that the Green function decays as $t\rightarrow -\infty$, and it is partially true. However as we can see
at not very large negative times the function develops a number of non-trivial features, - peaks which we can interpret as the ``most probable''
regions of times the particle can reach using the time machine. The corresponding results are present on Fig.\ref{fig:GreenFunctionDelta15}.
As an illustration we also provide a two dimensional plot for the leading $N=2$ winding at negative times, Fig\ref{fig:2DGreen}.

The origin of these peaks can be traced back to the fact that renormalized length
of a geodesic can be negative. Generically at small conformal weights $N=2$, $N=3$ and $N=4$ contributions are commensurate, but already at $\Delta \gtrsim 2$, higher entwinement terms are getting suppressed as compared to $N=2$. However at specific points, where $\cL_{ren}<0$, the corresponding
contributions to the Green function are getting enhanced in the large $\Delta$ limit\footnote{Strictly speaking, the geodesic approximation
is reliable only in this limit.}, forming a sharp peak. For instance, $N=2$ set of geodesics contains such a curve around $t_2=-1.9$, and $N=4$
set has a special point at $t_2\simeq -1.45$, Fig.\ref{fig:GreenFunctionDelta15}.

We have not performed numerical simulations for $N>4$, but we can not exclude that such negative length curves can appear also at large $N$.
The geodesic length is defined by lengths of internal segments (always positive) and lengths of the two boundary segments (that in principle
can be negative):
\be \cL^{ren}_{AB}=\cL^{ren}_{AO_1}+\cL^{ren}_{O_{2N}B}+\sum\limits_{i=1}^{N} \cL_{O_{2i-2}O_{2i-1}}\,.\ee
If $\cL^{ren}_{AO_1}+\cL^{ren}_{O_{2N}B}<0$, and $|\cL^{ren}_{AO_1}+\cL^{ren}_{O_{2N}B}|>\sum\limits_{i=1}^{N} \cL_{O_{2i-2}O_{2i-1}}$,
the Green function will get a contribution that does not vanish in the large $\Delta$ limit.
For a large number of internal segments it is not likely, but neither is impossible: while all internal lengths are
finite, the renormalized negative lengths might be of an arbitrarily huge absolute value:
\be L^{ren}_{AO_1}<0,\,\, |L^{ren}_{AO_1}|\gg 1\,, \ee
thus dominating over positive contributions.

In the case of a large conformal dimension it would mean that, if we were able to sum up contributions in all winding orders, the resulting Green function would have a shape of a comb with a number of peaks (in our calculations we discovered two of them). These peaks play a role of ``pit stops''
for a particle travelling in time, - they form a set of easily reachable coordinates in time. Hence we deal with specific ``negative time'' revivals.

\begin{figure}[t]
\begin{center}
\includegraphics[width=0.7\textwidth]{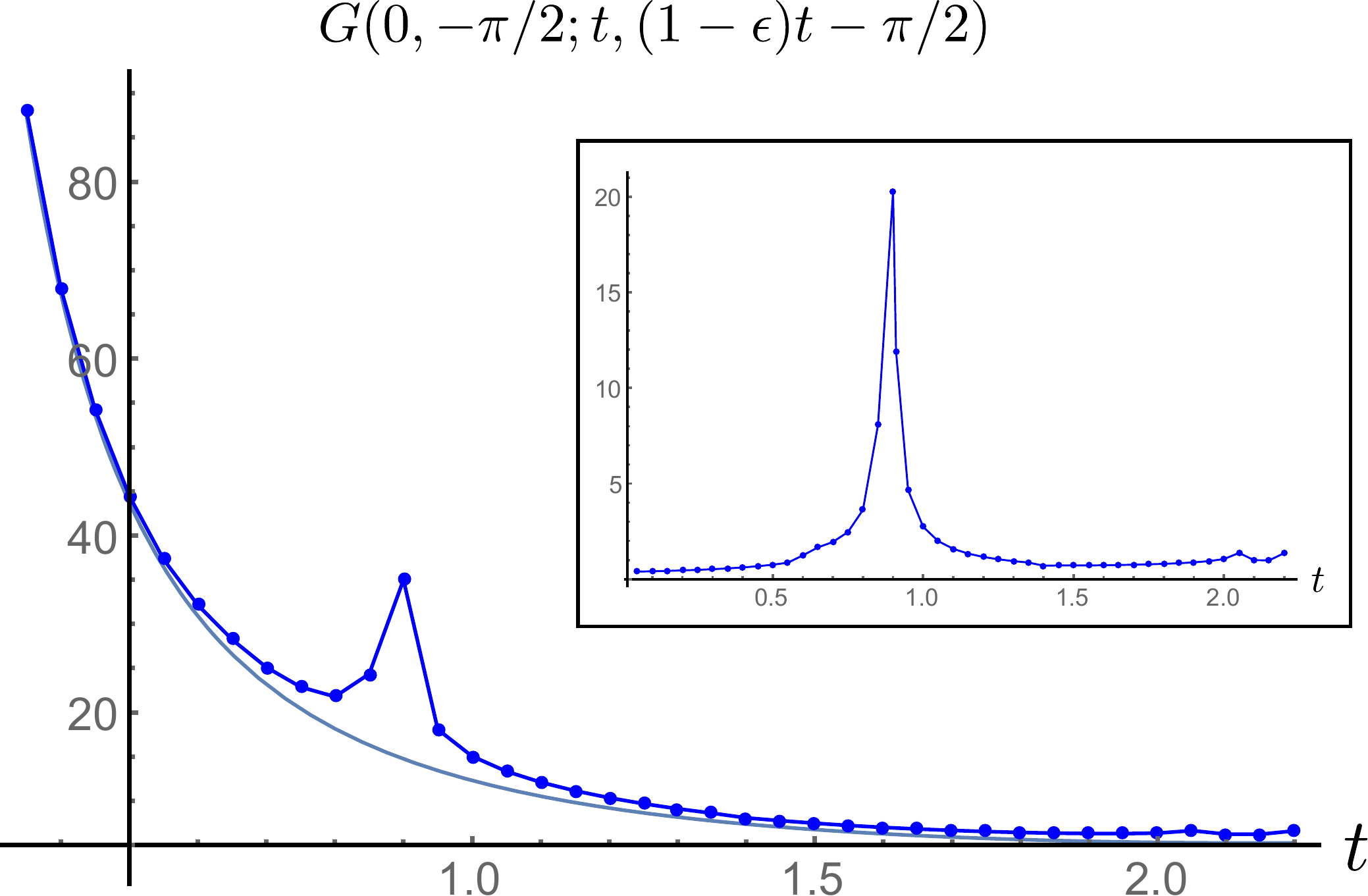}
\caption{The retarded Green function at conformal weight $\Delta=1$ at positive times. The plot demonstrates how the non-trivial $N=2$ windings
modify the original Green function ($N=1$ does not contribute when we consider the boundary theory within a single strip).
The offset plot represents the $N=2$ winding contribution separately.
We have made the Green's function timelike to avoid dealing with the light-cone singularity. Here $\epsilon=0.1$.}
\label{fig:PositiveTime}
\end{center}
\end{figure}

At positive times we have also discovered interesting features of the Green function. In the case of plain $AdS_3$ geometry
the dual light-like Green function (shifted away from the singularity) is decaying in time. In presence of the causality violating conical
defects we detected a new peak of a high weight, Fig.\ref{fig:PositiveTime}, signaling a revival of the excitation.

\section{Phases of the boundary field theory}
We have calculated the Green function numerically up to $N=4$ entwinements for the time machine geometry with
$\alpha=\sqrt{3}\pi$ and $\psi=1$. However it would be interesting to study how the properties of the Green function change
upon changing the strength and rapidities of the conical defects. We constructed the leading order $N=2$ contribution to the retarded Green function
at negative times for $\alpha \in \left(1.1\pi,...1.95\pi \right)$, and $\psi \in \left(0.1,...1.5 \right)$ with stepping
$\Delta \alpha =0.05 \pi $, $\Delta \psi = 0.05.$, paying special attention to the location in time and and strength of the revival peak.

\begin{figure}[t]
\begin{center}
\begin{tabular}{cc}
\includegraphics[width=0.5\textwidth]{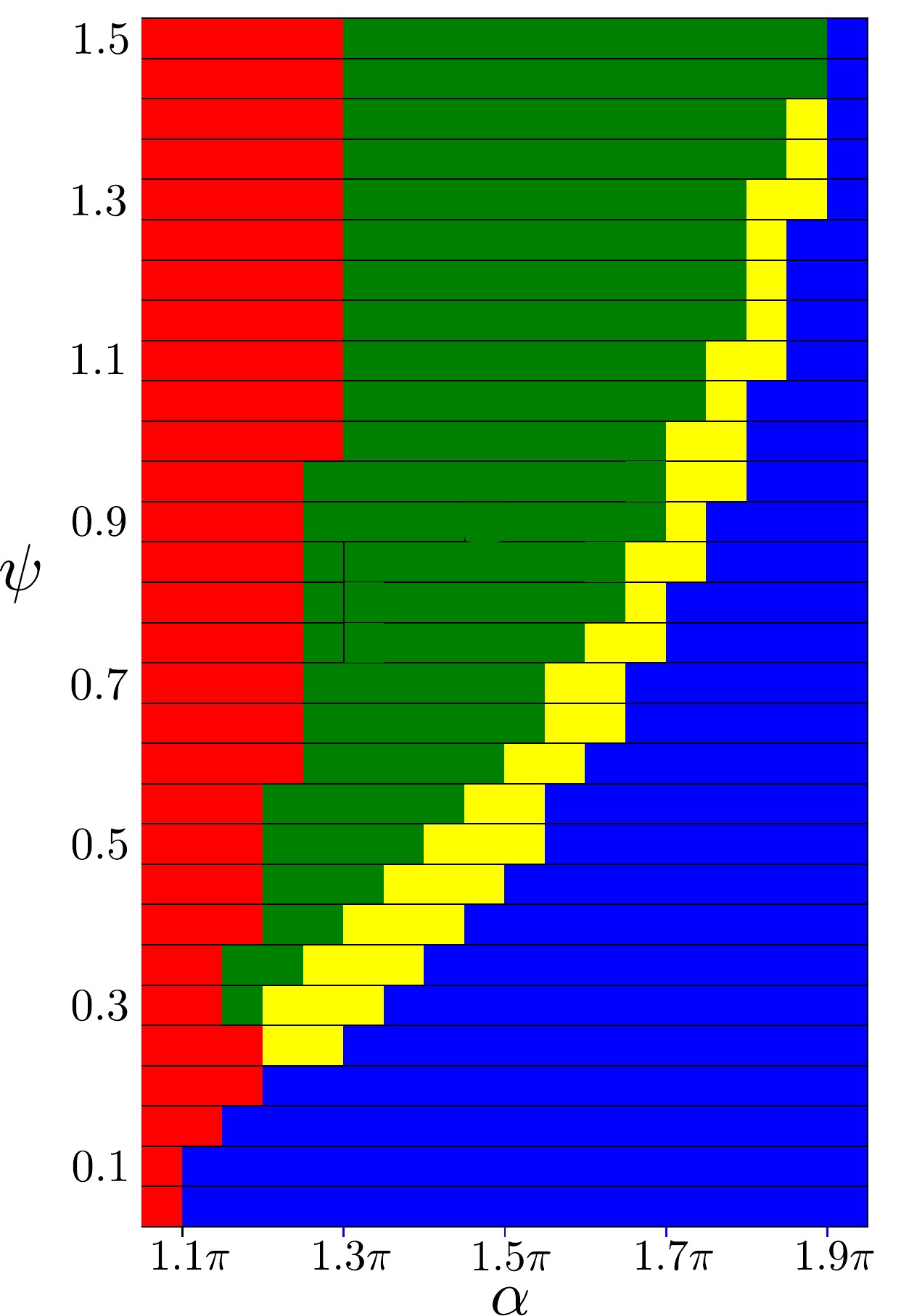}
\end{tabular}
\caption{The four different ``phases'' of the boundary field theory. Blue spots depict the region of forbidden geometries,
yellow spots are for the phase of negative time revivals, green spots form the region where the retarded Green function is non-zero
at negative times, but does not exhibit reviving peaks in the large $\Delta$ limit. Red spots are where at the leading $N=2$ order the boundary
field theory retarded Green function does not demonstrate causality violation (i.e. $G^{cr}(0,-\pi/2;t,-\pi/2)\equiv 0,\,\,\,t<0$) despite the presence of
the CTC in the bulk.
Everything is based on the numerical simulations of the leading
$N=2$ contribution to the retarded Green function. We expect higher order corrections to change the diagram qualitatively, but not quantitavely.}
\label{fig:PhaseDiagram}
\end{center}
\end{figure}

The results can be schematically summarized in a form of a phase diagram, Fig.\ref{fig:PhaseDiagram}:
\begin{itemize}
 \item If for a given value of $\alpha$ the rapidity $\psi$ is not large enough to prevent the system from collapsing; the DeDeo-Gott geometry
is forbidden (blue).
\item If for a given value of $\alpha$ the rapidity $\psi$ allows for the existence of the DeDeo-Gott time machine, but still not very large,
we clearly see the effect of revival, and the peak is sharper the closer $\psi$ is to the lower bound (yellow).
\item If the rapidity is too large, the causality is violated, but excitations just decay and do not revive at negative times anymore (green).
\item At very small values of $\alpha$ the retarded Green function does not exhibit any non-trivial features at negative time even in presence
of the closed timelike curves (red). However, this feature is likely just an artifact of $N=2$ approximation, and we do not expect it to
be there for higher windings.
\end{itemize}

\begin{figure}[t]
\begin{center}
\begin{tabular}{cc}
\subfigure[]{\includegraphics[width=0.5\textwidth]{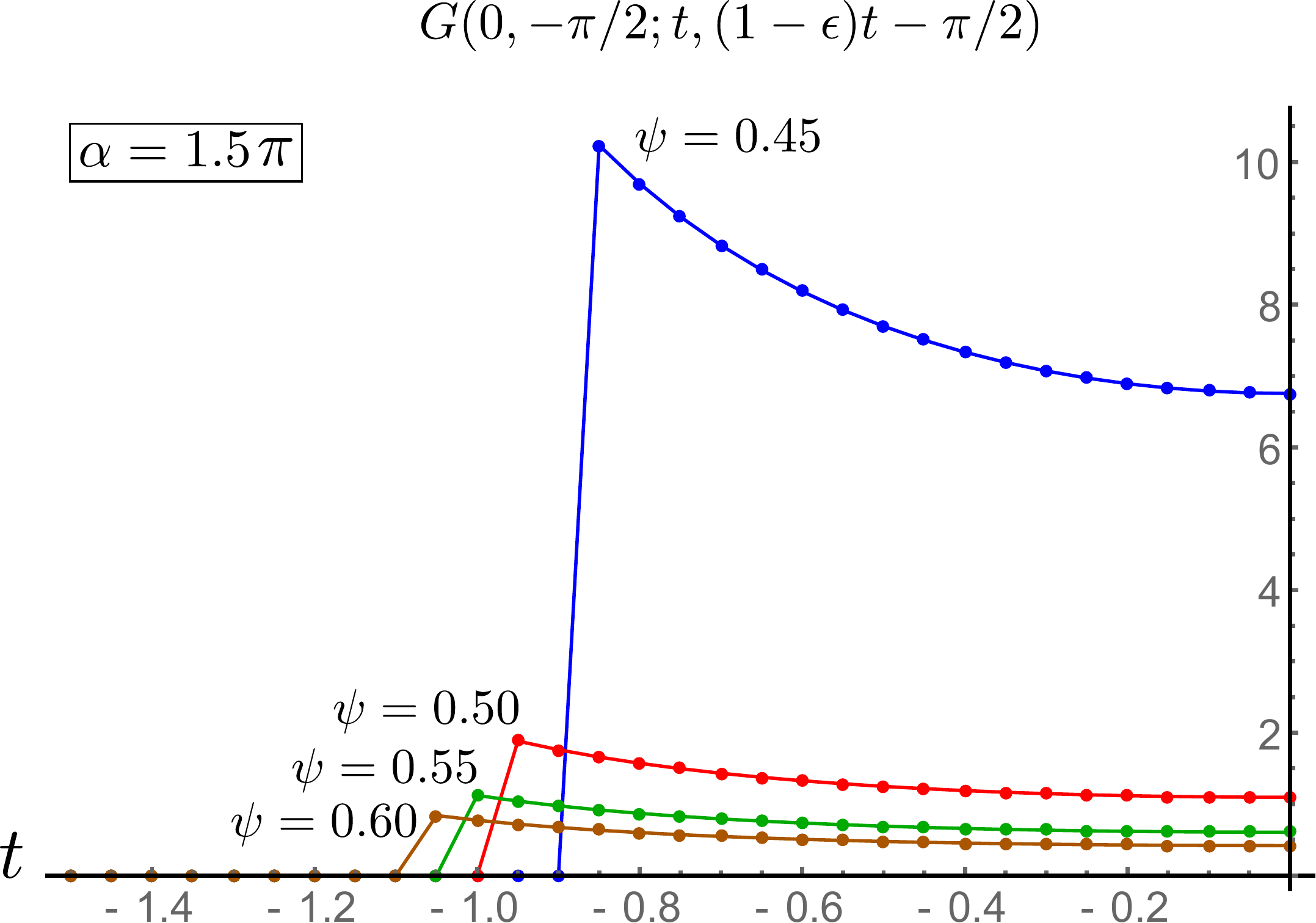}}
\subfigure[]{\includegraphics[width=0.5\textwidth]{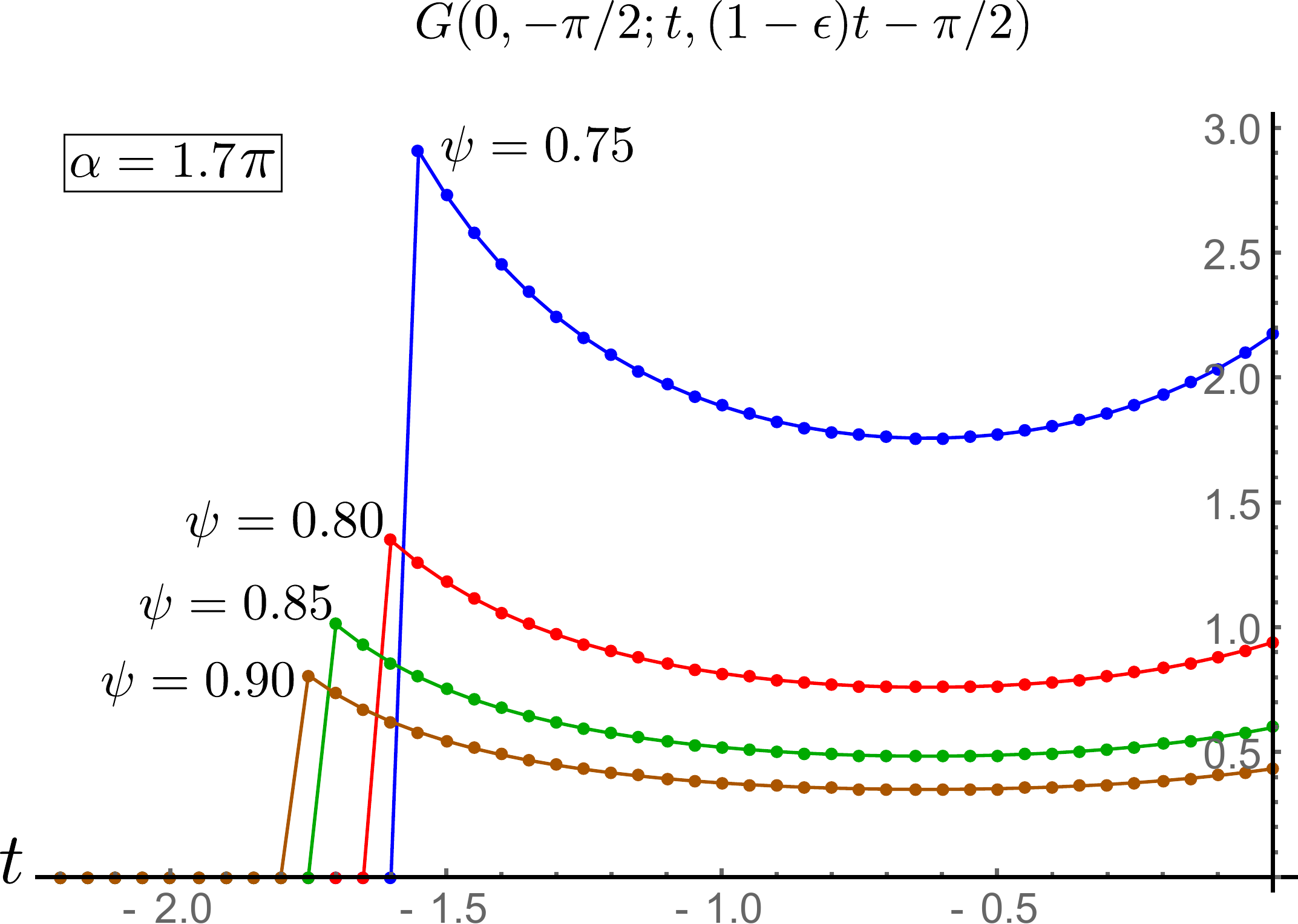}}
\end{tabular}
\caption{The negative time $N=2$ contribution to the retarded Green's function at (a) $\alpha=1.5$, $\psi=0.45,\,0.50,\,0.55,\,0.60$,
and (b) $\alpha=1.7$, $\psi=0.75,\,0.80,\,0.85,\,0.90$. Here $\Delta=1$. On each of the plots the first three peaks are getting stronger and sharper as $\Delta \rightarrow \infty$
(the yellow region on the phase diagram), while the fourth one is getting suppressed
in the same limit (the green region), so we do not consider it to be an actual revival of a non-causal excitation. Counterintuitively, the weight
of the Green function drastically decreases when we increase $\alpha$ or/and $\psi$.}
\label{fig:Green1517}
\end{center}
\end{figure}

The profiles of the Green's function at negative times are presented on Fig.\ref{fig:Green1517}(a,b)
for $\alpha=1.5\pi$ and $\alpha=1.7\pi$ respectively.
The fact that revivals are seen only at not very large rapidities (and the effect is stronger as closer $\psi$ to its
minimal possible value) is surprising and contrasts to how causality is broken in the bulk.
The stucture of CTC is defined by $\alpha$ and $\psi$, and the time jumps become stronger as the angle defects and rapidities are increased.
Thus we rather should expect that for high $\psi$
the time travelling along the CTC is more efficient in the sense that amplitudes of the {\it classical} free Green's function defined on the boundary
are getting enhanced as $\alpha$ $\psi$ grows. In the interacting holographic dual field theory the retarded Green's function is damped for larger $\alpha$
and $\psi$, so we can claim that causality in the boundary field theory is broken mildly as compared to the bulk.

Another interesting feature of this system is that while the overall weight of the Green function drastically decreases when
the rapidity $\psi$ is taken away from the ``forbidden region'' on the diagram, the actual past time penetration depth (i.e. the deepest
reachable point at negative times where $G^{cr}(0,-\pi/2;t,-\pi/2)\neq 0$) increases (though very moderately), and this is in agreement with the
``naive'' intuition.

\section{Conclusions}
In this paper we have analyzed properties of a two-point Green function in a (1+1)-dimensional field theory dual
to the DeDeo-Gott time machine geometry. Using the geodesic approximation we have shown that $AdS/CFT$ is capable
of describing a quantum field theory when causality is violated, and have shown that the corresponding boundary propagator has
remarkable features. We discovered that in presence of closed timelike curves in the $AdS$ bulk a causal propagation of an excitation
from the future to the past is possible on the boundary, and the retarded Green function exhibits peaks at certain negative times.
At positive times analytic structure of the Green function also changes, and new singularities arise.

Surprisingly, we have found that as we increase the strength of the conical defects $\alpha$ and the rapidity $\psi$, the causality violation
in the dual field theory is getting milder in the sense that the weight of the retarded Green's function at negative times decreases.

Contra to the previous results on the dynamics of physical systems in time machine backgrounds
\cite{VolovichGroshev,Echeverria:1991nk,Boulware,Friedman:1992jc} our calculations
have demonstrated that sometimes it is possible to define evolution of an interacting theory in a time machine without imposing
any additional self-consistency constraints. Despite the explicit non-causality the Green function does not have any
uncontrollable pathologies.

Our considerations leave a number of open questions. First of all, we have to understand how to interpret the boundary state dual to the DeDeo-Gott geometry, - whether this quantum state is pathological or just exotic yet physical state.
From the boundary point of view a single conical defect, if its angular deficit is $\alpha=2\pi(1-1/N)$, can be thought of as state created by a non-local twist operator in a conformal field theory \cite{Balasubramanian:2014sra}. But what it means to have such an interplay of two independently boosted defects
has to be clarified.

Another thing we have not touched on in the paper is the entanglement structure of the boundary state. We focused on the properties of the retarded
Green function, and thus analyzed the timelike quasigeodesics. However, even below the $\alpha=\pi$ threshold, when the CTC are not present in the system,
due to the lensing it is possible to connect timelike separated boundary points just by standard continuous spacelike geodesics. If a certain generalization
of the Ryu-Takayanagi conjecture \cite{Ryu:2006ef} is true in this case, it would mean the boundary state is timelike entangled \cite{Olson:2010jy}.
Possible physical outcomes of this fact is an interesting direction for future research.


\section{Acknowledgements}

We are grateful to Dmitry Ageev for collaboration during a certain stage of the project.
It is our pleasure to thank Simon DeDeo, Mikhail Katsnelson, Joris Vanhoof, and Jan Zaanen for illuminating discussions.
The work of I.A. is partially supported by RFBR grant 14-01-00707. A.B. and K.S. are supported in part by
the Netherlands Organization
for Scientific Research/Ministry of Science and Education
(NWO/OCW) and by the Foundation for Research
into Fundamental Matter (FOM). K.S. is also supported by a VICI grant of the Netherlands Organization for Scientific
Research (NWO).

\end{document}